\documentclass[]{aa}
\usepackage[varg]{txfonts}
\usepackage{graphicx}
\usepackage{booktabs}
\usepackage{amsmath}
\usepackage{colortbl}
\usepackage{subfig}
\usepackage{rotating}
\usepackage{amssymb}
\usepackage{latexsym}
\usepackage{ifthen}
\usepackage{enumitem}
\usepackage{hyperref}
\hypersetup{
     colorlinks   = true,
     citecolor    = blue
}
\usepackage{ulem}
\def\lsim{\lower.5ex\hbox{$\; \buildrel < \over \sim \;$}}
\def\gsim{\lower.5ex\hbox{$\; \buildrel > \over \sim \;$}}
\def\be{\begin{equation}}
\def\bea{\begin{eqnarray}}
\def\eea{\end{eqnarray}}
\def\ee{\end{equation}}

\def\bc{\begin{center}}
\def\ec{\end{center}}

\def\etal{{et al.}}
\def\ie{{$i.e.$,~}}
\def\ep{{e^--p^+}}
\def\el{{e^--e^+}}

\def\mbh{M_{\rm BH}}
\def\rg{r_{\rm g}}

\def\rsh{r_{\rm sh}}
\def\ep{{{\rm p}^+-{\rm e}^-}}

\def\msol{M_\odot}

\def\medd{\dot{M}_{\small{\rm Edd}}}
\def\betah{\beta_{\rm d}}

\newcommand\tab[1][1cm]{\hspace*{#1}}
\newcommand{\mdotin}{{\dot{\mathcal{M}}}_{\rm{in}}}
\newcommand{\rin}{r_{\rm in}}
\newcommand{\rci}{r_{\rm ci}}
\newcommand{\rcm}{r_{\rm cm}}
\newcommand{\rco}{r_{\rm co}}
\newcommand{\rc}{r_{\rm c}}

\newcommand{\vin}{v_{\rm in}}
\newcommand{\tp}{T_{\mbox{{\scriptsize p}}}}
\newcommand{\te}{T_{\mbox{{\scriptsize e}}}}
\newcommand{\tpi}{T_{\mbox{{\scriptsize {pin}}}}}
\newcommand{\tei}{T_{\mbox{{\scriptsize {ein}}}}}
\newcommand{\tpim}{T_{\mbox{{\scriptsize {pin|max}}}}}
\newcommand{\mdotinm}{{\dot{\mathcal{M}}}_{\rm{in|max}}}

\newcommand{\dqp}{\Delta Q_{\rm p}}
\newcommand{\dqe}{\Delta Q_{\rm e}}
\newcommand{\thetap}{\Theta_{\rm p}}
\newcommand{\thetae}{\Theta_{\rm e}}
\newcommand{\gamp}{\Gamma_{\mbox{{\scriptsize p}}}}
\newcommand{\game}{\Gamma_{\mbox{{\scriptsize e}}}}
\newcommand{\gamep}{Q_{\mbox{{\scriptsize ep}}}}
\newcommand{\polyp}{N_{\mbox{{\scriptsize p}}}}
\newcommand{\polye}{N_{\mbox{{\scriptsize e}}}}
\newcommand{\qpp}{Q^+_{\rm p}}
\newcommand{\qpm}{Q^-_{\rm p}}
\newcommand{\qep}{Q^+_{\rm e}}
\newcommand{\qem}{Q^-_{\rm e}}
\newcommand{\qbr}{Q_{\rm br}}
\newcommand{\qsy}{Q_{\rm syn}}
\newcommand{\qib}{Q_{\rm ib}}
\newcommand{\qic}{Q_{\rm ic}}
\newcommand{\qbrd}{\bar{Q}_{\rm br}}
\newcommand{\qsyd}{\bar{Q}_{\rm syn}}

\newcommand{\qicd}{\bar{Q}_{\rm ic}}

\newcommand{\thetapin}{\Theta_{\mbox{{\scriptsize p}}\scriptsize {\rm in}}}
\newcommand{\thetaein}{\Theta_{\mbox{{\scriptsize e}}\scriptsize{\rm in}}}

\newcommand{\nh}{n_{\rm in}}
\newcommand{\neh}{n_{\rm e in}}
\newcommand{\nph}{n_{\rm p in}}
\newcommand{\feh}{f_{\rm e in}}
\newcommand{\fph}{f_{\rm p in}}
\newcommand{\theh}{\Theta_{\rm e in}}
\newcommand{\thph}{\Theta_{\rm p in}}

\newcommand{\nel}{n_{\rm e}}
\newcommand{\np}{n_{\rm p}}
\newcommand{\neld}{\bar{n}_{\rm e}}
\newcommand{\npd}{\bar{n}_{\rm p}}

\newcommand{\me}{m_{\rm e}}
\newcommand{\mpr}{m_{\rm p}}
\newcommand{\gammainc}{\Gamma_{\rm inc}}

\newcommand{\betad}{\beta_{\rm d}}

\begin{document}

\title{Two temperature solutions and emergent spectra from relativistic accretion discs around black holes} 

\author{Shilpa Sarkar
		\inst{\ref{i1}, \ref{i2}}\thanks{E-mail: shilpa@aries.res.in}
		\and 
	Indranil Chattopadhyay
		\inst{\ref{i1}}\thanks{E-mail: indra@aries.res.in}	
		\and
		 Philippe Laurent
                \inst{\ref{i3}, \ref{i4}}\thanks{E-mail: philippe.laurent@cea.fr}
		}
\institute{Aryabhatta Research Institute of Observational Sciences 
(ARIES), Manora Peak, Nainital-263002, India\label{i1}
\and
Pt. Ravishankar Shukla University, Great Eastern Rd, Amanaka, Raipur, Chhattisgarh 492010 \label{i2}
\and
IRFU / Service d'Astrophysique, Bat. 709 Orme des Merisiers, CEA Saclay, 91191 Gif-sur-Yvette Cedex France\label{i3}
\and
Laboratoire Astroparticule et Cosmologie, Bâtiment Condorcet, 10, rue Alice Domont et Léonie Duquet, 75205 Paris Cedex 13 France\label{i4}
}

\date{Received -- / Accepted }

\abstract {} 
{We investigate two-temperature advective transonic accretion disc around a black hole and analyze its spectrum, in the presence of
radiative processes like bremsstrahlung, synchrotron and inverse-Comptonization. We would like to link the emergent spectra
with constants of motion of the accretion disc fluid. However, the number of unknowns in
two-temperature theory is more than the number of equations, for a given set of constants of motion. We intend to remove
the degeneracy using a general methodology and obtain a
unique solution and its spectrum.}
{
--
We use the hydrodynamic equations (continuity, momentum and energy conservation equation) to obtain
sonic points and solutions. To solve these equations of motion we use $4^{\rm {th}}$ order Runge-Kutta method. For spectral analysis,
general and special relativistic effects were taken into consideration. The system is, however degenerate. The degeneracy 
is removed by choosing the solution with maximum entropy, as is dictated by the second law of thermodynamics.}
{
--
A unique transonic solution exists for a given set of constants of motion. The entropy expression is a tool to select between the
degenerate solutions. We found that Coulomb coupling is a weak energy exchange term, which
allows protons and electrons to settle down into two different temperatures, hence justifying our study of two-temperature flows.
The information of the electron flow allows us to model the spectra. We show that the spectra of accretion solutions, depend on the
associated constants
of motion. At low accretion rates bremsstrahlung is important.
{A fraction of the bremsstrahlung photons may be of higher energy than the neighbouring electrons
hence energising them through the process of Compton scattering.}
Synchrotron emission, on the other hand, provides soft photons, which can be inverse-Comptonized to produce a hard power law
part in the spectrum. Luminosity increases with the increase in accretion rate of the system, as well as with the increase in BH mass.
However, the radiative
efficiency of the flow {has almost no dependence on} the BH mass, but {it} sharply rises with the increase in accretion rate. The spectral index, however, hardens with the increase in accretion rate, while it does not change much with the variation in BH mass. In addition to the
constants
of motion, value of plasma beta parameter and magnitude of magnetic dissipation in the system, also helps in shaping the spectrum.
Shocked solution exists in two-temperature accretion flows, {in a limited} region {of} the parameter space. It is
found that, a shocked solution is always brighter than a solution without a shock.
} 
{
-- 
An accreting system in two-temperature regime, admits multiple solutions for the same set of constants of motion, producing widely
different
spectra. {Comparing observed spectrum with that derived from a randomly chosen accretion solution, will give us a wrong
estimation of the accretion parameters of the system.} 
{The} form of
entropy measure {obtained by us}, have helped in removing the degeneracy of the solutions, allowing us to understand the physics of the
system, shorn of arbitrary assumptions. In this work, we have shown how the spectra and luminosities of an accreting system depends on the
constants of motion, producing solutions
ranging from radiatively inefficient flows to luminous flows. Increase in BH mass quantitatively changes the system, makes the
system more luminous and the spectral bandwidth also increases. Higher BH mass system spans from radio to gamma-rays. However, increasing
the accretion rate around a BH of certain mass, has little influence in the frequency range of the spectra. 
}
\maketitle
\titlerunning{Relativistic two temperature accretion solution} 

\section[Introduction]{Introduction}
\label{sec1}

{Accretion is the {primary mechanism which could explain the observations of radiation coming from
microquasars and active galactic nuclei (AGN). Stellar mass black holes (BH) and neutron stars are thought to reside in microquasars
and super-massive BHs reside in AGNs.}
The energy released due to accretion of matter onto {relativistic} compact objects, like
neutron stars and BHs, is {a fraction} of the rest mass energy of the matter falling onto it. }
The study of accretion process was initiated
by \citeauthor{hl39} in 1939,
{where} they studied accretion of matter onto a Newtonian star passing through the interstellar medium (ISM). 
In 1952, \citeauthor{b52} gave
the first full analytical solution for spherical flows (also known as Bondi flows) around a static star. This theory is 
also relevant for
the case of stellar winds. {He} showed that accretion/wind is characterized by a unique transonic solution having maximum
entropy.
{But it took 10 years, until the discovery of quasars and X-ray
sources in 1960's, that accretion phenomena gained popularity.
\citet{s64} and \citet{z64} extensively investigated accretion as a probable mechanism for driving
and powering these luminous objects. It was concluded that the Bondi accretion model produced luminosities which is too low to
explain the observations. Matter being radially falling, have short infall timescales compared to
their cooling
timescales, leading to low radiative efficiency of such flows}. This led to the development of the famous \citeauthor{ss73} disc
model (SSD) or the Keplerian disc
model
{\citep{pr72,ss73,nt73}}. In this model it was assumed that matter is rotating in Keplerian orbits having negligible radial
velocity {and} pressure gradient terms. 
The shear between differentially rotating matter gives rise to viscous {stresses}  which helps
in removing angular momentum
outwards, {allowing matter to} spiral inwards finally falling onto
the central object. The
time taken for inspiral, allows the matter to emit for a longer duration, unlike spherical flows.
{Since at every radius the angular momentum of the disc is Keplerian, therefore the disc has to be geometrically thin. In other words, it implies that the heat generated due to viscous dissipation needs to be efficiently radiated away such that the angular momentum distribution
remains close to the Keplerian one \citep[hence the name `cool discs'][]{king12}.}
In SSDs, matter and radiation are in thermal equilibrium, where each annulus of the disc emits a
blackbody spectrum {(or, depending on opacity, a modified version of it)} peaked at the {temperature of the annulus of the
disc}.
The composite spectrum is hence the sum of these blackbodies 
and is called the modified multicoloured blackbody spectrum.
Although this model could successfully regenerate
the thermal part of the spectrum from sources associated with BHs,
but was unable to explain the non-thermal part. In addition, the assumption of Keplerian angular velocity
at each annulus implied that the disc is arbitrarily terminated at the inner stable circular orbit or ISCO.
{Soon it was also concluded that these discs were thermally and secularly unstable \citep{prp73, le74,art96}.} In 1975,
\citeauthor{tp75} argued that the instability present
at the inner region of SSD could expand into an optically thin gas-pressure dominated region.
\citet[hereafter SLE76]{sle76} considered
this {geometrically thick and optically thin} puffed up region {to be composed of protons and electrons described
by two different temperature distributions}.
{Using this model, they successfully} reproduced the hard component part of Cygnus X-1 spectrum from $8-500~keV$.
But unfortunately, SLE model {too}, was found to be thermally unstable \citep{p78}. If the disc is heated,
{it expands reducing its number density and thereby its cooling rate.
This makes the system even hotter leading to a runaway thermal instability}. Although unstable, this paper served as one of the
cornerstones in the two-temperature accretion theory.

The models that have been discussed hitherto, suffered from simplifying assumptions, for example, the cooling rate
{at each radius of the disc was equated with the
heating rate} and the advection term was not properly dealt with. In general, the heating and cooling rates need
not be equal and some part of the heat could be advected inwards along with the bulk motion of the system.
In 1988, \citeauthor{aetal88} extensively investigated advection, in their "slim" optically thick accretion disc model and found
that the solutions obtained were thermally and viscously stable.
Importance of advection was further demonstrated using self similar solutions in the works of \citet{ny94,aetal95}.
These discs are today broadly classified as advection-dominated accretion flows or ADAFs and
\citet{i77} was the first to propose it \citep[for review also see][]{bl97,bl01}.

A general conclusion can be drawn from the above models that an accretion flow need not be Keplerian throughout but could also be
sub-Keplerian or a combination of both. Also the flow has to be transonic. Matter very far away from the horizon is subsonic
whereas BH boundary condition insists that matter should cross the horizon at the speed of light. Thus accreting matter has to pass through
atleast
one sonic point, {or in other words, BH accretion solution is necessarily transonic. \cite{b52} in his seminal paper
have already highlighted the importance of transonicity for spherical accretion flows. In 1980, Liang and Thompson (hereafter, \citet{lt80}) argued that similar to spherical flows which is characterized by a single sonic point, rotating flows around BHs are characterized by multiple sonic points. 
\citet{f87} extended their work and presented in details the nature of sonic points in transonic
rotating accretion flows. He concluded that even though a BH does not possess any hard surface, it can undergo a shock transition
in the presence of multiple sonic points \citep[also see][]{c89}.} 

{All the works mentioned above (except SLE76), assumed one-temperature accretion flows. One-temperature flows
are based on the fact that the timescale of the energy exchange process (like Coulomb coupling)
between the ions and electrons is shorter or comparable
to the dynamical time scale of the system,
allowing the system to effectively settle down into a single temperature distribution \citep {lb05,cc11,kc14}.
But it is to be noted that
in many astrophysical cases, infall timescales are in general much shorter than Coulomb collision timescales, i. e., Coulomb
coupling between the protons and electrons are not strong, allowing the two species to equilibrate to two different temperature
distributions \citep{rees82,ny95}.
Therefore, in addition to the advective transonic nature of an accretion flow around a BH, the gas is likely, to be in the
two-temperature regime. 
Also, the electrons are more prone to radiative cooling, compared to ions.
This makes the electron temperature deviate largely from the protons especially in the
inner regions of the accretion disc. 
}

{After the seminal paper of SLE76, two-temperature assumption was largely used to model accretion flows as it could
successfully reproduce the observed spectrum, electrons being the main radiators. \citet{cmt84} included advection and solved the energy equation
(or, first law of thermodynamics)
for spherical accretion flows (\ie with no angular momentum), and
assumed free fall velocity field with radial dependence of the form: $v \propto r^{-1/2}$. Transonic nature of the flow was
not taken into account, but emission processes 
relative to protons and electrons, were discussed briefly}. {In 1995, \citeauthor{ny95} (hereafter NY95) incorporated
angular
momentum and extensively discussed the nature of two-temperature, optically thin accretion discs.
The equations of motion were solved
under the self-similar assumption.} 
It is to be noted that, self-similarity in accretion solution around BH, is
plausible only at a large distance from the horizon and not near it. Additionally, a self similar solution is not
transonic. 
NY95 neglected the electron advection term and assumed that the heating and cooling rates of
electrons {to be} equal. {There have been other notable works done in two-temperature regime, where although
the issue of transonicity was bypassed, but radiative transfer part was properly treated. One such work was by
\citet{ct95}, where the spectrum was computed assuming two-components in the accretion flow: a Keplerian and another sub-Keplerian.
Exact Comptonization model was incorporated, following which they discussed the variation in spectrum with the change in mass of BH and
accretion rate. \cite{mc05}, went a step further and included a non-thermal distribution of electrons along with the general
thermal distribution, inside the accretion flow.
Similar to \cite{cmt84}, the velocity field was assumed to be free fall, but shocked solution and its signature in the observed
spectrum was qualitatively studied.}

In 1996, \citeauthor{netal96}, obtained the first global transonic, two-temperature accretion solutions,
{where the advection terms were considered without any simplifying assumption. 
However at the outer boundary, the authors assumed that ion
temperature to be a fraction of the virial temperature and Coulomb coupling was equated with bremsstrahlung. 
\citet{metal97} extended \citeauthor{netal96}'s work to calculate the
spectrum.
In addition, they slightly modified the outer boundary conditions, where the total gas temperature (and not the ion temperature) was
assumed to
be a fraction of the virial temperature and equated
Coulomb coupling with the total cooling of electrons (bremsstrahlung, synchrotron and Comptonization). }
Recently, a more general transonic advective two temperature accretion solution was obtained by \citet{rb10},
where the viscous stress was assumed to be proportional to the sum of ram pressure and gas pressure
\citep[see][for details on this particular viscosity prescription]{c96}.
In this work, limited class of solutions were studied. 
This work was extended and global class of transonic two-temperature solutions including accretion-shock, were
investigated by \citet{detal17}. {The most striking feature of the transonic two-temperature works mentioned above is that, the solutions depend on the choice of inner or outer boundary conditions, but from single-temperature
hydrodynamics, we know that transonic solutions are unique for a given set of constants of motion. We highlight this issue in greater
details below.}

Hydrodynamic equations, even in the single temperature regime, admit infinite number of solutions, but a transonic solution 
is physically favoured because it has the highest entropy among all possible global solutions \citep{b52}. {In addition,
the location of the sonic point corresponds to a unique boundary condition.}
The number of hydrodynamic equations for flows in one-temperature and two-temperature regime, are exactly the same,
but there is one more variable in case of two-temperature flows (\ie the existence of different ion and electron temperature
distribution). 
Obtaining a self-consistent two-temperature flow would require solving the basic hydrodynamic equations,
but since there is one more
flow variable in the two-temperature regime, the system is degenerate. We obtain a large number of transonic solutions,
for the same set of constants of motion of the flow. Hint of this {problem of degeneracy} was reported briefly in LT80,
however, the problem was skirted by parameterizing the temperatures of proton and electron to some constant
value, arguing that the
coupling between these species is unknown. {As previously mentioned,}  few authors also assumed an arbitrary proton or
electron temperature at the boundary where they
started integrating, to find solutions. A global treatment of two-temperature problem would require all the equations to be solved
self-consistently without taking recourse to any set of arbitrary assumptions on temperature values at any boundary. This problem
of degeneracy was
identified and a prescription to obtain a unique transonic two-temperature solution was reported in Sarkar \& Chattopadhyay
(2019, hereafter \citet{sc19}).
But it was applied to flows having zero angular momentum. Spherical flows
have single sonic points which simplifies our problem, allowing us to focus only on the issue of degeneracy
in two-temperature regime. In \citet{sc19}, we reported that, for a given set of constants of motion, infinite transonic solutions
exist, each having a sonic point
property different from the rest. The question that arises is, which solution to choose, since nature does not prefer degeneracy.

In one-temperature regime, the transonic solution is unique { \citep{lb05,bdl08}}, but for reasons cited above,
two temperature solution is degenerate
for the same set of constants of motion. 
Following \citet{b52}, we can look for highest entropy solution, in order to remove degeneracy. However
the two-temperature energy equation, in adiabatic limit, is not integrable
\citep[unlike in one-temperature regime,][]{kscc13,kc13,kc14,ck16}.
This prevented us from obtaining an analytical expression for measure of entropy in the two-temperature regime. The integration of
the energy equation is spoiled by the presence of Coulomb coupling term.
But remembering the fact that near the horizon gravity overpowers any other interaction or processes, we can neglect the
Coulomb coupling term.
Thus we reported in \citet{sc19}, for the first time, a form of entropy measure, which could only be applied near the horizon.
{Using the formula obtained,}
we measured the entropies at the horizon for the transonic spherical two-temperature solutions obtained for a given set of constants of
motion. We saw that the entropy maximized at a certain solution. From the second law
of thermodynamics we select the solution with maximum entropy as the solution which nature would prefer. This solved the problem of
degeneracy in two-temperature model. It may however be noted, that spherical flows were simple to handle owing to the presence of
single sonic points. 

{It may be noted that, like the flow speed, the thermal state of the flow} around a BH is {also}
trans-relativistic in nature. That is
very far away from the horizon, matter is thermally
non-relativistic and as it approaches the BH, it could be sub-relativistic or relativistic. Matter is referred as thermally
relativistic
when its thermal energy is comparable to or greater than its rest mass energy ($kT/mc^2 \gsim 1$, where $k=$ Boltzmann constant,
$T=$ temperature, $m=$ mass of the species) and adiabatic index ($\Gamma$) $\sim 4/3$ and is non-relativistic if
its thermal energy is less than the rest mass energy ($kT/mc^2 < 1$) and $\Gamma\sim 5/3$. So, not only the temperature but
also the mass
of the {constituent particles  decides the relativistic nature of the flow}.
So, in two-temperature flows, where we consider two species with masses
differing by $\gtrsim 1000$ times, an equation of state (EoS), with fixed adiabatic index, is untenable. \citet{c08,cr09},
proposed an
approximate EoS for multispecies flow (CR EoS, hereafter), which is analytical and computationally easy to handle. Although
approximate,
it matches perfectly well \citep{vkmc15} with the relativistically perfect EoS, which was obtained by
\citet{c38}. 
Later the authors extended the use of CR EoS for dissipative, relativistic accretion disc too \citep{ck16}.
To incorporate the trans-relativistic nature of protons and electrons, one need to utilize a variable adiabatic index EoS.

In this paper we would like to extend our previous \citet{sc19} model to rotating flows, {and thereby, remove the
degeneracy of accretion solutions around BHs. Needless to say, choosing any one of the degenerate solutions,
would give us an incorrect picture of the accretion parameters of the BH.}
{To handle the trans-relativistic nature of these flows, we use the CR EoS
which freed us from specifying the adiabatic indices of each species.}
Hydrostatic equilibrium  along the disc thickness is maintained at each radius of the
disc. 
Radiative processes with proper special
and general relativistic corrections has been incorporated, to compute the spectrum.
{
In this paper,
we would like to analyse how the correct unique accretion solution depend on the constants of motion of the flow, along with the
mass of the BH and discuss how these properties play an active role in shaping the spectrum.
Further, we intend to study the relative contribution of various radiative processes, on the emitted spectrum and also investigate which part of the disc is likely to contribute the most in the emitted spectrum.
In addition, we want to check whether an accretion shock imparts any special spectral signature. 
Apart from these, we would like to study the
dependence of radiative efficiency and spectral index on the variation in accretion rate of the BH and its mass and also with
the presence of different radiative processes.}

This paper is divided into the following sections. In Sect. \ref{sec2}, we will give a brief overview of the basic equations and assumptions
used to model the flow. In Sect. \ref{sec:solproc} we will discuss the solution procedure to find a unique transonic two-temperature solution.
We will show the results in Sect. \ref{sec:result} and finally conclude in Sect. \ref{sec4}.

\section{The two temperature advective disc model : assumptions and governing equations}
\label{sec2}

In this paper, our main intention is to obtain all possible accretion solutions onto a BH
and compute the typical spectrum corresponding to each mode of accretion. 
Solution of a rotating accretion flow is much more complicated than spherical accretion, due to the presence of multiple
sonic points. Therefore, the spectra is different for different kinds of solutions. Furthermore, two-temperature flows are degenerate,
which makes the method to obtain unique solution very important. 

In the coming subsections we will discuss the equations that we have used to model two-temperature accretion flow. We will also give an overview of the radiative processes incorporated and the methodology to implement these processes in curved space-time.

\subsection{Equations of motion (EoM)}
The space-time metric around a non-rotating BH is described using a Schwarzschild metric :
\begin{equation}
ds^2=g_{tt} c^2dt^2+
g_{rr}dr^2
+g_{\theta \theta}d{\theta}^2+g_{\phi \phi}d\phi^2,
\end{equation}
where the metric tensors are expressed as, $-g_{tt}=g_{rr}^{-1}=(1-2G\mbh/c^2r)$ and $g_{\theta \theta}=g_{\phi \phi}=r^2$,
since an accretion flow is described around the equatorial plane.
Here, $r,~ \theta$ and $\phi$ are the usual spherical coordinates and $t$ is the time coordinate, $G=$ Gravitational constant and
$\mbh=$ mass of BH.
It is to be noted that throughout the paper, we have employed a system of units where the unit of length, velocity and time are 
defined as
$\rg=G\mbh/c^2,~c$ and $\rg/c=G\mbh/c^3$ respectively.
All the variables used in the rest of the paper have been written in this unit system unless mentioned otherwise. The BH system
modelled is in steady state and is axis-symmetric, therefore $\partial/\partial t=\partial/\partial \phi=0$. Moreover, at any radius we 
assume that only the radial gradient of any quantity is dominant, therefore, 
$\partial/\partial \theta=0$.

Radial component of the momentum balance equation is:
\begin{equation}
    u^r \frac{du^r}{dr}+\frac{1}{r^2}-(r-3)u^\phi u^\phi+(g^{rr}+u^r u^r) \frac{1}{e+p} \frac{dp}{dr}=0,
\label{eq:rad-ns}
\end{equation}
where $e$ and $p$ are the internal energy density and isotropic gas pressure respectively, measured in the local fluid
frame and $u^\mu$s are the components of four-velocity.
The mass accretion rate is obtained by integrating the conservation of four mass-flux:
\begin{equation}
    \dot{M}=4\pi \rho H u^r r,
	\label{eq:accretion-rate}
\end{equation}
where, $\rho=n(\mpr+\me)=$ local mass density of the flow, {$n$ is the particle number density}, $\mpr$ and $\me$ are the mass of
proton and electron respectively, and
$H$ is the local half-height of the disc. Here, ${\dot M}$--- the accretion rate, is a constant of motion throughout the flow.
Half-height is calculated assuming hydrostatic equilibrium along the vertical direction of the disc \citep{lasota1994,cc11} which
can be written as,
\begin{equation}
    H=\sqrt{\frac{pr^3}{\rho\gamma_\phi^2}}.
	\label{eq:halfheight}
\end{equation}
Also, from the fact $u_\mu u^\mu=-1$, we obtain:
\begin{displaymath}
   -u_t=\sqrt{\left(1-\frac{2}{r}\right)}\gamma_v \gamma_\phi,
	\label{eq:ut}
\end{displaymath}
where, $\gamma_v$ and $\gamma_\phi$ are the Lorentz factors in the radial and azimuthal directions
respectively and are defined as
$\gamma_v=\sqrt{1/(1-v^2)}$ and $\gamma_\phi=\sqrt{1/(1-v_\phi^2)}$ where $v_{\phi}=\sqrt{-u_\phi u^\phi /u_tu^t}$ and
$v$ is the velocity in the local co-rotating frame. 
It can be shown that $v^2=\gamma_{\phi}^2v_{\hat{r}}^2$, where,
$v_{\hat{r}}=\sqrt{-u_ru^r/u_tu^t}$. The total {Lorentz} factor therefore is, $\gamma=\gamma_v \gamma_\phi$.\\\\
The first law of thermodynamics or the energy balance equation is $u_\mu T^{\mu\nu}_{;\nu}=\Delta Q$ and can be written as :
\begin{equation}
    u^r \left[ \left( \frac{e+p}{\rho} \right) \rho_{,r} -e_{, r} \right]=\Delta Q.
	\label{eq:flt}
\end{equation}
Here, $\Delta Q=Q^+-Q^-$ where, $Q^+$ and $Q^-$ represents the rate of heating and cooling present in the
flow for all the species. 
These rates are in units of ergs cm$^{-3}$ s$^{-1}$, which are {converted into} geometric units before being used
in the above equation. 

Coulomb coupling serves as an energy exchange process, which transfers energy between protons and
electrons. In the single temperature case, Coulomb
coupling being infinitely strong, protons and electrons equilibrate locally into a single temperature distribution
\citep{lmc98,lrc11,lckhr16,ck16}. But for
two-temperature flows, it is not strong enough, hence allowing protons and electrons to thermalise
at two different temperatures. In other words, the timescale for protons and electrons to attain a thermal equilibrium and settle down
into a single temperature is more than the timescale in which, each of the two populations thermalise separately. 
To describe such flows, we need to use two separate energy equations, {one} for protons and {another for} electrons. These two
energy equations are not independent, and as discussed, coupled by the Coulomb coupling term
which acts as a cooling term for protons and a heating term for electrons, if the proton temperature is higher than electron
temperature.

If we integrate equation of motion (Eq.~\ref{eq:rad-ns}) with the help of
{energy} equation (Eq.~\ref{eq:flt}),
we obtain the generalized Bernoulli constant and is given by,
\begin{equation}
    E=-hu_t \textrm{exp}(X_f),
\label{eq:gbc}
\end{equation}
where, $h=(e+p)/\rho=$ specific enthalpy and $X_f=\int \frac{{\dqp}+{\dqe}}{\rho h u^r}dr$. Here, $\Delta Q_i=Q_i^+-Q_i^-$,
represents the difference in the heating and cooling
rates of the $i^{\rm{th}}$ species. The generalised Bernoulli constant is conserved all throughout the flow, even in the presence of heating and cooling. $X_f$ term mainly arises due to the presence of dissipation.
In case of adiabatic flows, with no dissipation, $X_f=0$ and $E \rightarrow {\cal E}=-hu_t$, which is the canonical form of 
relativistic Bernoulli constant \citep{lppt75,cc11}.
\subsection{EoS and the final form of the EoM}
It is to be noted that In this subsection barred variables have been used to denote dimensional quantities and unbarred variables
are, as before, non-dimensional. In order to solve the above equations of motion we need to
supply an equation of state (EoS). In this paper we have used the Chattopadhyay-Ryu (CR) EoS given by \citet{c08,cr09}.
The form of CR EoS for multispecies flow is :
\begin{equation}
    \bar{e}=\sum_{i} \bar{e}_i=\sum_{i} \left[ \bar{n}_im_ic^2 +\bar{p}_i \left( \frac{9\bar{p}_i+3\bar{n}_im_ic^2}{3\bar{p}_i+2\bar{n}_im_ic^2} \right) \right],
\label{eq:eos}
\end{equation}
where, the summation is over $i^{\rm{th}}$ species. Since, in this paper, we have  considered the
accretion flow to be composed of
protons and electrons ($\ep$) only, so $i$ represents these two species. Number density $(n)$, mass density $(\rho)$ and isotropic
gas pressure $(p)$, present in Eq.~\ref{eq:eos} can be
represented in dimensional form in the following way :
\begin{equation}
    \bar{n}=\sum_{i} \bar{n}_i=\bar{n}_p+\bar{n}_e=2\bar{n}_e,   
    	\label{eq:nd}
\end{equation}
\begin{equation}
{\bar \rho=\sum_{i} \bar n_i m_i= \bar{n}_{\rm e}\me+\bar{n}_{\rm p} \mpr=\bar{n}_{\rm e} \me \left( 1+\frac{1}{\eta}\right)=\bar{n}_{\rm e} \me \tilde{K},}
	\label{eq:md}
\end{equation}
\begin{equation}
\bar p=\sum_{i}  \bar p_i=\sum_{i} \bar n_ik {T}_i=\bar{n}_{\rm e} k(  {{T}}_{\rm e}+{{T}}_{\rm p})=\bar{n}_{\rm e} \me c^2\left(\thetae+\frac{\thetap}{\eta} \right),
	\label{eq:pressure}
\end{equation} 
\begin{displaymath}
\textrm{where, $\eta=\me /\mpr$ and $\tilde{K}=1+1/\eta$. ${T}_i$ is the temperature in units of kelvin, while }\Theta_i=\frac{kT_i}{m_ic^2} \textrm{ is the non-dimensional temperature}
\end{displaymath}
defined w.r.t the rest-mass energy of the respective $i^{\rm th}$ species.
The EoS, Eq.~\ref{eq:eos} can be simplified using Eqs.~\ref{eq:nd}-\ref{eq:pressure} to,
\begin{equation}
{\bar e}=\bar{n}_{\rm e} \me c^2\left(f_{\rm e}+\frac{f_{\rm p}}{\eta} \right)=\frac{\bar \rho c^2f}{\tilde{K}},
	\label{eq:eos-sim}
\end{equation}
\begin{displaymath}
\textrm{where,~~}f_{i}=1+\Theta_i\left(\frac{9\Theta_i+3}{3\Theta_i+2}\right) \textrm{~~ and ~~}
f=f_{\rm e}+\frac{f_{\rm p}}{\eta}.
\end{displaymath}
Polytropic index and adiabatic index can be written as,
\begin{equation}
{N_i}=\frac{df_i}{d\Theta_i}	\tab \mbox{and} \tab \Gamma_i=1+\frac{1}{N_i}.
\end{equation}
The equation for half-height (Eq.~\ref{eq:halfheight}), can be simplified to :
\begin{equation}
H=\sqrt{\frac{[r^3-\lambda^2(r-2)]}{\tilde{K}}\left( \thetae+\frac{\thetap}{\eta}\right)}
	\label{eq:halfheight-2}
\end{equation}
{Here, $\lambda=-u_\phi/u_t$ is the specific angular momentum of the flow. Angular momentum plays a very important role in
accretion disc physics, because it can significantly modify the infall time scale.}.
Using Eqs.~\ref{eq:nd}-\ref{eq:halfheight-2}, we can simplify energy equation (Eq.~\ref{eq:flt}) and obtain two
differential equations for temperature, one for proton and another for electron. They are as follows:
\begin{equation}
\frac{d\thetap}{dr}=-\frac{2\thetap}{2\polyp+1}\left({\cal{A}}+ \frac{1}{v(1-v^2)}\frac{dv}{dr}\right) - \mathbb{P}\eta,
\label{eq:dtpdr}
\end{equation}
\begin{equation}
    \frac{d\thetae}{dr}=-\frac{2\thetae}{2\polye+1}\left({\cal{A}}+ \frac{1}{v(1-v^2)}\frac{dv}{dr}\right) - \mathbb{E},
	\label{eq:dtedr}
\end{equation}
respectively, where, 
\begin{displaymath}
{\cal{A}}=-\frac{r}{r(r-2)}-\frac{3r^2-\lambda^2}{2[r^3-\lambda^2(r-2)]}~~,~~~~
\mathbb{P}=\frac{2\Delta Q_{\rm{p}}\tilde{K}}{\rho u^r (2\polyp+1)},\textrm{~~~~and~~~~~~}
\mathbb{E}=\frac{2\Delta Q_{\rm{e}}\tilde{K}}{\rho u^r (2\polye+1)}.\\
\end{displaymath}
If we simplify the radial component of the relativistic momentum balance equation (Eq.~\ref{eq:rad-ns}) using Eqs.~\ref{eq:nd}-\ref{eq:dtedr}, we get the expression of gradient of three-velocity, which has the form :
\begin{equation}
    \frac{dv}{dr}=\frac{\cal{N}}{\cal{D}},
	\label{eq:dvdr}
\end{equation}
\begin{displaymath}
\textrm{where,~~} {\cal{N}}=-\frac{1}{r(r-2)}+\frac{\lambda^2\gamma_\phi^2(r-3)}{r^4} +a^2 {\cal{A}}+\frac{\gamp \polyp \mathbb{P}+\game \polye \mathbb{E}}{h\tilde{K}}-\frac{\Delta Q}{\rho h u^r} \textrm{~~~~~~and~~}
\quad \mathcal{D}=\frac{v}{1-v^2}\left( 1- \frac{a^2}{v^2}\right).\\
\end{displaymath}
The effective sound speed ($a$) has been defined
as 
\begin{equation}
a^2=\frac{\mathcal{G}}{h\tilde{K}},
\end{equation}
\begin{displaymath}
\textrm{where,~~}{\cal{G}}~=~\frac{2\gamp \polyp \thetap}{\eta(2\polyp+1)}+\frac{2\game \polye\thetae}{(2\polye+1)}.
\end{displaymath}

\subsection{Heating and cooling processes included in the flow }
In the coming subsections, we briefly discuss the processes which leads to heating and cooling of the plasma in the accretion flow.
\subsubsection{Heating due to magnetic dissipation}
\label{sec:magdiss}
Magnetic field in the medium surrounding the BH would be frozen into the highly conductive infalling plasma. As the matter falls
inwards,
its magnetic field strength would increase by $1/r^2$ and magnetic energy density ($B^2/8\pi$) by $1/r^4$.
In 1971, \citeauthor{s71} argued that before the magnetic energy density exceeds the thermal energy density, turbulence and
hydromagnetic instabilities would lead to reconnection of magnetic field lines. In other words, it means that the magnetic energy
density is limited by
equipartition with thermal energy density. This magnetic energy dissipated would heat up the matter, either protons or electrons or
both, ensuring relativistic temperatures even far away from the BH. The expression for dissipative heating rate as given by
\cite{ip82} ,
\begin{equation}
\bar{Q}_{\rm B}=\frac{3u^r c}{2r\rg}\frac{B^2}{8\pi} = \frac{3u^r c}{2r\rg} \betad \bar{p}=
\frac{3u^r c}{2r\rg} \betad \bar{n}_{\rm{e}}k(\tp+\te)\quad \rm{{ergs~ cm^{-3} s^{-1}}},
\label{eq:magheat} 
\end{equation}
The above equation is a measure
of the heating due to magnetic dissipation. However, there are uncertainties in the estimates of heating processes which is controlled
by $\betad$. We have used $\betad=0.001$, unless stated otherwise, {as a representative case}.
{In Sect. \ref{sec:betad}, we have varied the value of $\betad$ and studied how the solutions depend on it.} 
In this work we assume that both protons and electrons can absorb the magnetic energy dissipated. Hence, we can write,
\begin{equation}
\qpp=\delta Q_{\rm B} \tab\mbox{and}\tab \qep=(1-\delta) Q_{\rm B}
\end{equation}
where, $\delta$ is the uncertainty parameter, which dictates the amount of heat absorbed by protons, the rest being absorbed by
electrons. There is insufficient knowledge present in literature discussing this issue. Thus, throughout our work, for simplicity,
we consider $\delta=0.5$, which means that 50\% of this heat would
go to protons and the rest 50\% into electrons, unless otherwise mentioned.

In our present study, we have investigated inviscid flows, since proper handling of {general relativistic} (GR) form of viscosity
in transonic flows is not trivial. The shear tensor in GR contains derivative
of $u_\phi$, $v$ and other terms \citep[Peitz \& Appl 1997, hereafter][]{pa97}. It is impossible to obtain a solution if all
the terms of the shear tensor is considered. \citet{pa97} proposed an approximate form of shear tensor by neglecting
all derivatives of $v$ and then presented a limited class of solutions. \citet{ck16} used the same form of viscosity
but obtained the full range of solutions. Also they computed mass-loss from such advective accretion solutions.
We envisage that the method to obtain viscous solution is not easy, since the Bernoulli parameter of viscous flow
has no analytical form. 
In addition, the sonic point is not known apriori and needs to be obtained as a part of eigenvalue of the solution. 
Moreover, the angular momentum on the horizon needs to be computed. And yet the solutions obtained are limited, because
the viscosity is still phenomenological and various terms of
the relativistic version of the shear
tensor has to be neglected in order to obtain a solution.
Two-temperature regime further complicates the problem, as has been pointed out above. Most of the
works done in literature assumed Newtonian form of viscosity
or the \citeauthor{ss73} $\alpha$-viscosity prescription (hereafter SS), use of which, in our GR model would be inappropriate.
So we avoided the use
of any form of viscosity since the prime focus of this paper, is to present a novel methodology to obtain unique transonic
two-temperature solutions for accretion discs around BHs. 
In addition, it has been extensively shown in Figs. 2h-2i; 3h-3i of \citet{ck16} that the specific angular momentum ($\lambda=-u_\phi/u_t$) and bulk angular momentum
($L=hu_\phi$) in the last few $100 \rg$, is almost constant
and sub-Keplerian. This is to be expected, as gravity
supersedes all other interactions near the BH horizon. In order to exhibit the qualitative effect of viscosity, as a representative
case, one-temperature, viscous accretion
disc solutions are presented in Appendix \ref{app:1}, by following the methodology of \citet{ck16}. 
We have used two forms of viscosity, where the viscous stress tensor is given by (i) $t_{r \phi}= -2 \eta_{\rm vis} \sigma_{r \phi} $, 
(abbreviated as PA) and (ii)
$t_{r \phi}=-\alpha_{\rm vis} p$ (SS form of viscosity).
The form of $\sigma_{r\phi}$ of PA is adopted from \citet{pa97,ck16}, but presently, we have assumed the
dynamical viscosity coefficient $\eta_{\rm vis}=\rho h \nu_{\rm vis}$, instead
of $\eta_{\rm vis}=\rho \nu_{\rm vis}$, where $\nu_{\rm vis}$ is the kinematic viscosity. 
In Fig. \ref{fig:14}a, we show that for both the form of viscosities (PA and SS),
$\lambda \approx$ constant for $r\lsim 1000\rg$. 
In Fig. \ref{fig:14}b, we plot the heat dissipated by various processes.
PA form of viscosity is stronger than the SS type of viscosity
while it is in general much weaker than $Q_{\rm B}$. It is comparable to
$Q_{\rm B}$ only in a very narrow region.
%
%
Since close to the horizon, angular momentum variation of the flow is quite small and the viscous
heat dissipated is less than the magnetic heating,
as a result, for simplicity, in this paper we studied accretion in the weak viscosity limit. 
We concentrated on obtaining a self-consistent two-temperature, transonic,
rotating accretion solutions, by considering low angular momentum flows at the outer boundary and compute the spectra
for such flows.
We study how accretion rate, angular momentum and mass of the central black hole might affect the solution 
as well as emergent spectra from such solutions.

\subsubsection{Coulomb coupling} 
As discussed before, Coulomb coupling ($\gamep$) serves as an energy exchange process between protons and electrons.
Therefore,
\begin{equation}
\qpm=\qep=\gamep
\end{equation}
The expression for Coulomb coupling in cgs units ($\rm{ergs~cm^{-3}s^{-1}}$) is given by \cite{sg83},
\begin{align}
& \bar{Q}_{\rm{ep}} = \frac{3}{2}\frac{\me}{\mpr}\neld \npd  \sigma_T c k \frac{\tp-\te}{K_2 \left(1/\thetae\right) K_2 
\left( {1/\thetap}\right)} \textrm{ln } \Lambda_c  \left[ \frac{2(\thetae+\thetap)^2+1}{\thetae+\thetap} K_1 
\left( \frac{\thetae+\thetap}{\thetae \thetap}\right)+ 2K_0 \left( \frac{\thetae+\thetap}{\thetae\thetap}\right)  \right],
\label{eq:cc}
\end{align}
where, $\sigma_T$ is the Thomson scattering cross-section, $K_{\rm{i}}(x)$'s are the modified Bessel functions of second kind and i$^{\rm th}$ order
and ln $\Lambda_c$ is the Coulomb logarithm which is set equal to 20.

%
{There have been apprehensions that more efficient energy exchange processes might exist between ions and electrons, in addition to Coulomb
coupling. In that case the accretion flow may settle down into a single temperature
distribution \citep{phinney81}. \citet{bc88} used plasma waves and \citet{sharma07} used magneto-rotational instability to increase the energy
exchange between the two-species inside the flow. 
However, some authors have raised doubts 
about the effectiveness of these processes \citep{ob14,af13}. In this paper, we have ignored any type of collective effects and
considered only Coulomb coupling as the main energy exchange process between the protons and electrons. 
}

\subsubsection{Inverse bremsstrahlung} 
Inverse bremsstrahlung ($\bar Q_{\rm{ib}}$) is a radiative loss term for the protons, the expression of which is given below
\citep{boldtser1969}:
\begin{equation}
   {\bar{Q}_{\rm{ib}} =1.4 \times 10^{-27} \neld^2 \sqrt{\frac{\me}{\mpr}\tp}.} \tab \rm{{ergs~ cm^{-3} s^{-1}}}.
	\label{eq:ib}
\end{equation}

\subsubsection{Radiative mechanisms leading to cooling of electrons}
The cooling of electrons could be caused by three basic cooling mechanisms (1) bremsstrahlung ($\qbr$), (2) synchrotron ($\qsy$)
and (3) inverse-Comptonization ($\qic$).
Emissivity due to bremsstrahlung (in $\rm{{ergs~ cm^{-3} s^{-1}}}$) is given by \citet{nt73},
\begin{equation}
    \qbrd=1.4 \times 10^{-27} \neld^2 \sqrt{\te}\left(1+4.4 \times 10^{-10}\te\right).
	\label{eq:brem}
\end{equation}
We have used thermal synchrotron radiation in our model, following the prescription of \citet{wz00}. The emissivity is given by:
\begin{equation}
  \qsyd=\frac{2 \pi}{3} \frac{\nu_t^3}{r r_g} \me \thetae,
	\label{eq:sync}
\end{equation}
where, $\nu_t$ is the turnover frequency, above which the plasma is optically thin to synchrotron radiation and below which
it is highly self-absorbed by the electrons itself. For calculation of $\nu_t$ we need the information of magnetic field in the flow.
For this purpose we have considered a stochastic magnetic field which is in partial or total equipartition with the gas pressure,
same as mentioned before in Sect. \ref{sec:magdiss}. Thus, $B=\sqrt{8\pi\beta \bar{p}}$. We set $\beta = 0.01$ throughout this
paper unless otherwise mentioned. {In Sect. \ref{sec:betasync}, we have varied the value of $\beta$ and have discussed how the spectrum depends on it.} \\
The soft photons generated through thermal synchrotron process could be inverse-Comptonized by the electrons present in the plasma.
It is given by \citep{wz00},
\begin{equation}
 \qicd={\zeta}\qsyd,
	\label{eq:comp}
\end{equation}
where, $\zeta$ is the enhancement factor. It is expressed as,
\begin{displaymath}
\zeta=3{{\varphi}} \left( \frac{x_t}{\thetae}\right)^{\alpha_0-1}
\left[ \gammainc \left( 1-\alpha_0, \frac{x_t}{\thetae} \right)+\frac{6\gammainc(\alpha_0)P_{\rm{sc}}}{\gammainc(2\alpha_0+3)}\right].
\end{displaymath}
\begin{displaymath}
\textrm{Here, $\gammainc$ is the incomplete gamma function, }x_t=\frac{h\nu_t}{m_ec^2}, ~~\varphi=\frac{[1+(2\thetae)^2]}{[1+10(2\thetae)^2]}, ~~\textrm{$\alpha_0$ is the
spectral index which can be defined as :}
\end{displaymath}

\begin{equation}
   \alpha_0=-\frac{{\rm{ln}}~P_{\rm{sc}}}{{\rm{ln}}~A}
	\label{eq:alpha}
\end{equation}
It is the slope of the power law photons generated, due to inverse-Comptonization, at each radius. {Therefore, the net
spectral index ($\alpha$) of the final inverse-Compton spectrum is obtained from the contributions of all the
$\alpha_0$'s from each radius of the disc}.
In Eq. \ref{eq:alpha} $A=1+4\thetae+16\thetae^2$, is the average amplification factor in energy of photon per scattering and
$P_{\rm{sc}}=1-{\rm{exp}}(-\tau_{\rm{es}})$, is the probability that a photon is scattered.
$\tau_{\rm{es}}$ is defined as the optical depth of the medium where electron scattering is important, expression of which
is given by \citep{turolla1986}, 
\begin{equation}
\tau_\textrm{es}=0.4 \left[ 1+ \left(2.22 \te \times 10^{-9} \right) ^{0.86}\right]^{-1}\bar{\rho} H\rg.
\end{equation}
\subsubsection{Compton heating}
As have been discussed before, less energetic photons would cool the flow through the process of inverse-Comptonization.
But if the temperature of the electrons is less than the temperature of the photons present in the flow, then the electrons will gain
energy via Compton scattering. This would lead to Compton heating of the electrons.
We assume that this has the same expression as that of inverse-Comptonization, but the sign changes \citep{esin97}. It causes heating rather than cooling. Therefore,
\begin{equation}
\bar{Q}_{\rm{e}}^+=\bar{Q}_{\rm{comp}}
\end{equation}
\subsubsection{Final expressions for $\Delta Q_{\rm{p}}$ and $\Delta Q_{\rm{e}}$}
So, to conclude we have taken for heating and cooling of protons~:
\begin{equation}
\qpp=\delta{Q}_{\rm B} \tab \mbox{and} \tab \qpm=\gamep+\qib,
\end{equation}
respectively. And for heating and cooling of electrons~:
\begin{equation}
\qep=(1-\delta){Q}_{\rm B}+\gamep+Q_{\rm{comp}} \quad \mbox{and} \quad \qem=\qbr+\qsy+\qic
\end{equation}
respectively.\\ Therefore,
$\Delta Q_{\rm{p}}=\delta{Q}_{\rm B}-\gamep-\qib$ and
$\Delta Q_{\rm{e}}=(1-\delta){Q}_{\rm B}+\gamep+Q_{\rm{comp}}-\qbr-\qsy-\qic$\\

{In this paper, we have ignored pion production and its contribution to the observed spectra, as well as ignored pair production
arising
from the interactions of high energy photons present inside the disc. Later in Sect. \ref{sec:result},
we will show from posteriori calculations that the contribution of both the processes are not significant.}

\subsection{Entropy accretion rate expression}
\label{sec:entropy}
If we switch off the explicit heating and cooling of protons and electrons, the gradient of proton and electron temperatures
becomes (using Eq.~\ref{eq:flt}):
\begin{equation}
\frac{d\thetap}{dr}=\frac{\thetap}{\polyp}\frac{1}{\np}\frac{d\np}{dr} + \frac{\gamep \eta \tilde{K}}{\rho u^r \polyp} 
{\rm{and}},
\label{eq:fst1law2}
\end{equation}
\begin{equation}
 \frac{d\thetae}{dr}=\frac{\thetae}{\polye}\frac{1}{\nel}\frac{d\nel}{dr} -
 \frac{\gamep\tilde{K}}{\rho u^r \polye}.
\label{eq:fst1law22}
\end{equation}
Due to the presence of Coulomb interaction term, we cannot integrate the above equation and obtain an analytical form\footnote{In single temperature regime, absence of Coulomb coupling makes it easier to integrate the corresponding equation and obtain an
analytical measure of entropy for all $r$ \citep{kscc13}.}. Hence, we cannot have a measure of entropy at every point of the flow.

However, an analytical expression is admissible only in regions where $\gamep$ is negligible. Such a region is near the horizon
($\rin$), where gravity overpowers any other interaction.
The integrated form of Eqs.~\ref{eq:fst1law2} and \ref{eq:fst1law22} are:
\begin{align}
&\neh=\kappa_1 ~ {\rm exp}{\left({\frac{\feh}{\theh}}\right)}\theh^{\frac{3}{2}}(3\theh+2)^{\frac{3}{2}}\\
&\nph=\kappa_2 ~{\rm exp}{\left({\frac{\fph}{\thph}}\right)}\thph^{\frac{3}{2}}(3\thph+2)^{\frac{3}{2}},
\label{eq:ent_2}
\end{align}
where, $\kappa_1$ and $\kappa_2$ are the integration constants which are measures of entropy.
Neutrality of the plasma implies $\neh=\nph=\nh$. Subscript `${\rm in}$' indicates quantities measured just outside the horizon. Therefore,
\begin{equation}
\nh^2 =\neh \nph \Rightarrow \nh =\sqrt{\neh \nph}
\end{equation}
Thus, we can write, 
\begin{align}
\nh=\kappa \sqrt{{\rm exp}{\left({\frac{\feh}{\theh}}\right)}~{\rm exp}{\left({\frac{\fph}{\thph}}\right)}\theh^{\frac{3}{2}}\thph^{\frac{3}{2}}{(3\theh+2)^{\frac{3}{2}}}  (3\thph+2)^{\frac{3}{2}}},
\end{align}
\begin{displaymath}
\textrm{where, }\kappa=\sqrt{\kappa_1 \kappa_2}
\end{displaymath}
The expression of entropy accretion rate, using Eq.~\ref{eq:accretion-rate} can be written as,
\begin{align}
\nonumber \mdotin &=\frac{\dot{M}}{4\pi\kappa (\me+\mpr)}\\ 
\nonumber &=\frac{4\pi \nh (\me+\mpr) H_{\rm in} u^r_{\rm in}  \rin}{4\pi\kappa (\me+\mpr)}\\
&=\left [ \sqrt{{\rm exp}{\left({\frac{\feh}{\theh}}\right)}{\rm exp}{\left({\frac{\fph}{\thph}}\right)}\theh^{\frac{3}{2}}\thph^{\frac{3}{2}}{(3\theh+2)^{\frac{3}{2}}(3\thph+2)^{\frac{3}{2}}}} \right ] H_{\rm in} u^r_{\rm in}  \rin .
\label{eq:ear}
\end{align}
\subsection{Sonic point conditions}
\label{sec:spc}
The mathematical form of Eq.~\ref{eq:dvdr} suggests that, at some point of the flow, where $a=v$, the denominator (${\cal{D}}$) goes to $0$. Then, for the flow to be continuous, the numerator (${\cal{N}}$) also has to go to $0$. This is called the sonic point of the flow. Sonic points exist whenever $dv/dr={\cal{N}}/{\cal{D}}=0/0$. Thus, the sonic point conditions are:
\begin{equation}
-\frac{1}{r_{\rm{c}}(r_{\rm{c}}-2)}+\frac{\lambda_{\rm{c}}^2\gamma_{\phi_{\rm{c}}}^2(r_{\rm{c}}-3)}{r_{\rm{c}}^4} +a_{\rm{c}}^2 {\cal{A}}_{\rm{c}}+\frac{\Gamma_{\rm{pc}} N_{\rm{pc}} \mathbb{P}_{\rm{c}}+\Gamma_{\rm{ec}} N_{\rm{ec}} \mathbb{E}_{\rm{c}}}{h_{\rm{c}}\tilde{K}}-\frac{\Delta Q_{\rm{c}}}{\rho_{\rm{c}} h_{\rm{c}} u^r_{\rm{c}}}=0,
\label{eq:spc1}
\end{equation}
and,
\begin{equation}
\frac{v_{\rm{c}}}{1-v_{\rm{c}}^2}\left( 1- \frac{a_{\rm{c}}^2}{v_{\rm{c}}^2}\right)=0 \Rightarrow v_{\rm{c}}=a_{\rm{c}}.
\label{eq:spc2}
\end{equation}
Here, the subscript `$c$' corresponds to the value of flow variables at the sonic point. The derivative at the sonic point
$dv/dr \vert _c$, is computed using the L'Hospital rule.
\subsection{Shock conditions}
\label{sec:shockcond}
The relativistic shock conditions or the Rankine-Hugoniot conditions \citep{t48} are :\\
Conservation of mass flux across the shock : $[\dot{M}]=0$\\
Conservation of energy flux : $[\dot{E}]=0$\\
Conservation of momentum flux : $[\Sigma h \gamma_v^2 v^2 +W] = 0 $\\
where, $\Sigma=2\rho H$ and $W=2pH$
are the vertically averaged density and pressure respectively.
The square brackets denote the difference of the quantities across the shock.
\subsection{Observed spectrum}
\label{sec:spec}
In this model we have incorporated radiative processes like bremsstrahlung, synchrotron and inverse-Comptonization which
give rise to
emissions spanning over the whole electromagnetic spectrum. This emission
(measured in units of ergs s$^{-1}$ Hz$^{-1}$) when plotted as a function
of frequency (in units of Hz) gives us the spectrum. The spectrum is an observational tool that helps us in determining the intrinsic
properties of any object (distant or nearby). Thus, calculation of the correct spectrum is important. While obtaining a solution for a
given set of flow parameters, \ie mass of the BH, accretion rate etc, we have information of the emission coming from each radius, or
in other words, the spectrum at each radius is known,
which is a function of the local $v$, $\rho$ and $T$. When the contribution from each radius is added,
we get the total observed spectrum. The model presented in this paper is in the pure GR regime, so we take into account all the general
and special relativistic effects, in the observed spectrum. Below, we explain the methodology to compute the spectrum.

Let us assume that the isotropic emissivity per frequency interval per unit solid angle, in the fluid rest frame is $j_\nu$. 
If we transform this emissivity to a local flat frame, then by using special-relativistic transformations this becomes :
\begin{equation}
j{'}_{\nu '}=j_\nu \frac{1-v^2}{(1-v \mbox{ cos} \theta ')^2}~~~~~\mbox{and}~~~~ \nu{'}=\nu \frac{\sqrt{1-v^2}}{(1-v \mbox{ cos} \theta ')}
\end{equation}
Here, $\theta '$ is the angle which the velocity of the fluid element directed inwards makes with the line of sight.

It is to be noted that all the photons emerging from the disc need not reach the observer. Some would be captured by the BH due to its
extreme gravity. The amount of emission captured by the BH, depends on its distance from the BH. The expression to calculate this was
given by \cite{zn71} :
\begin{equation}
\lvert \mbox{ cos} \theta^*  \rvert = \sqrt{\frac{27}{4} \left( \frac{2}{r} \right) ^2 \left( \frac{2}{r} \right) +1 }
\end{equation}
where $\theta^*$ is the angle within which photons will be captured by the BH and hence lost.

Now if we integrate the emissivity expression over the whole volume of the disc and on all solid angles, taking into account
$\theta ^*$, we get the luminosity of the system as a function of frequency and hence the spectrum. We have also accounted for the
gravitational redshift which introduces a factor of $\sqrt{1-2/r}$ in the observed frequency. Furthermore, if we want to calculate the
bolometric luminosity of the system, we need to integrate the frequency dependent luminosity over all the frequencies. For more details 
on the calculation of spectrum, see \cite{shap73}. In the total spectrum, there are signatures of all the emission processes and 
has been discussed extensively in the results section. Bremsstrahlung emission always comes in the high frequency part of the spectrum.
When the accretion rate of the system is low, {the contribution from bremsstrahlung emission is visible in the spectrum}.
Synchrotron emission is characterized by the
turnover/absorption frequency ($\nu_t$). Inverse-Comptonization, on the other hand, is identified as a power law part in the spectrum,
following the relation $F_\nu \varpropto \nu^\alpha$, $\alpha$ being the spectral index.


\section{Solution Procedure} \label{sec:solproc}
Accretion disc around BHs are transonic in nature and {may} possess multiple sonic points \citep[LT80,][]{f87,c89}.
{The nature of the sonic point is also dictated by the slope of the solution at the sonic point.
If the slope (i. e., $dM/dr\vert_c$, here $M=v/a$ is the Mach number)
admits two real roots at the sonic point, then a solution can actually pass through it.
These type of sonic points are termed as X-type or
saddle-type. The accretion
solution corresponds to the negative slope while excretion solution corresponds to the positive slope. 
However, if both the roots of the slope are imaginary or complex, then the matter cannot pass through them.
These sonic points are called O-type (imaginary slope) and spiral-type (complex) respectively.}
{The combined effect of the flow parameters like $E,~\lambda,~\&~{\dot M}$, determines
the number of sonic points formed inside an accretion flow as well as topology of the solution. An accretion flow with
low values of $\lambda$ admits only one outer sonic point ($\rco$; located at larger distance from the BH), while
those with higher values of $\lambda$ admit only
inner sonic points ($\rci$; closer to the BH). In the intermediate $\lambda$ range,
accretion disc may admit a maximum of three sonic points : inner ($\rci$), middle ($\rcm$) and  outer ($\rco$),
where $\rci$ and $\rco$ are X-type while the middle sonic may be spiral or O-type depending on whether the system is
dissipative or non-dissipative respectively \citep[for further information on sonic points see,][]{h77,lt80,ftrt85,f87,c89}.
Flows with high value of $E$,
generally admit only one $\rci$. ${\dot M}$ on the other hand controls radiative cooling and modifies the thermal state of the
flow. This in turn modifies the range of $E$ and $\lambda$ which allows multiple sonic point formation.

{In Sect. \ref{sec:sonicptprocedure}, we describe the method to find sonic points and in Sect. \ref{sec:removdeg}, we show that the
transonic solutions are degenerate and discuss extensively how to remove the degeneracy.}

\subsection{Method to obtain sonic points {in} two-temperature flows}
\label{sec:sonicptprocedure}
\begin{figure}[h!!!!!!]
\begin{center}
 \includegraphics[width=17.cm,height=7cm,trim={0.2cm 11.cm 0.2cm 4.2cm},clip]{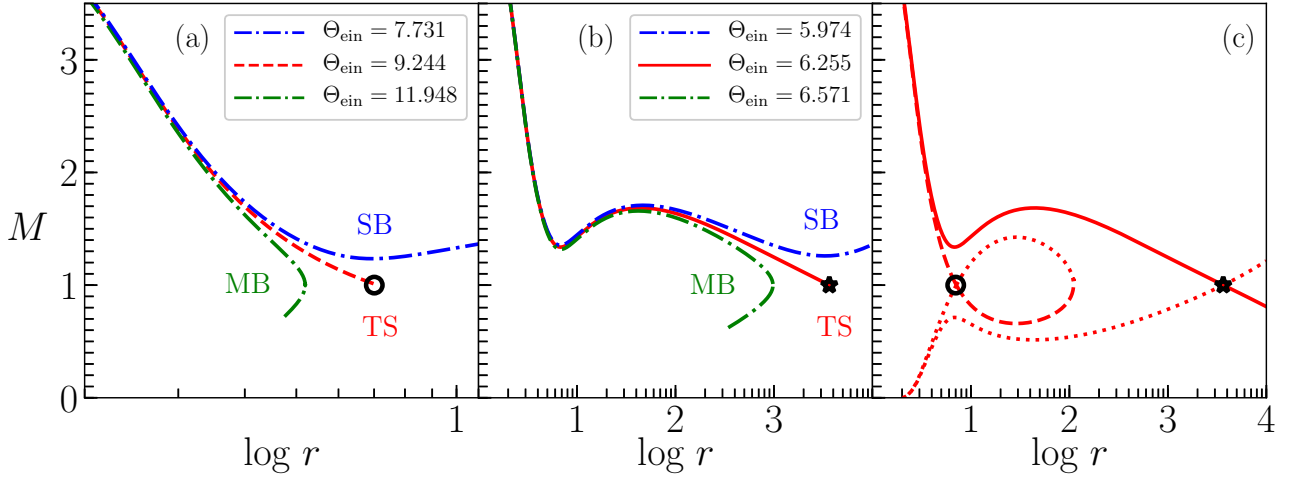}
  \caption{Method to find sonic points. Solutions are presented in terms of Mach number $M~(=v/a)$ vs log $r$ plot.
 $\thetapin=7.162\times 10^{-2}$ for all iterations. 
 Panel (a) iterations to obtain inner sonic point $\rci$ (black circle) and panel (b) iterations to obtain outer sonic point $\rco$ (black star). Various branches plotted are multivalued branch (MB; green dashed-dot),
 transonic (TS; red dashed) and supersonic (SB; blue dashed-dot).
 Respective $\thetaein$s are mentioned inside the panels.
 Panel (c) plots full set of transonic solutions Global accretion solution (red solid) through $\rco$ and accretion solution through
 $\rci$ (red, dashed) is plotted. Equatorial global wind (through $\rco$) and non global wind (through $\rci$) are represented
 using red dotted curve.
  The accretion disc flow parameters used are $\lambda=2.5$, $E=1.000045$, ${\dot M}=0.001\medd$ and $\mbh=10\msol$.}
\label{fig:1}
 \end{center}
\end{figure}
{In single temperature regime}, location of sonic point and its property, is unique for a given set of constants of motion.
In dissipative systems, sonic points
are not known a priori {and is obtained self-consistently by integrating the equations of motion}.
{So, presently in the two temperature regime we follow exactly the same procedure
to solve the equations, as is done in the
single temperature realm}. We need to select some fixed boundary from where we can start integrating $dv/dr$, $d\thetap/dr$ and
$d\thetae/dr$, to find the sonic point.
As $r\rightarrow$ $2\rg$, $v\rightarrow c$ and $E$ being a constant of motion, is also defined on the horizon. However,
one cannot start integration from $r=2\rg$ because of coordinate singularity on the horizon.
Therefore, we select a point asymptotically close to the horizon ($\rin=2.001\rg$).
It is here, where gravity overpowers any other processes or interactions, therefore infall timescales are much less than any other timescales.
In other words, at $r=\rin$, $X_f\rightarrow 0$ and $E\rightarrow {\cal{E}}=-hu_t$ (from Eq. \ref{eq:gbc}).
Simplifying this, we obtain an expression of $\vin$ in terms of $E,~\lambda,~\rin,~\thetapin$ and $\thetaein$.
{We list down the procedure to obtain a transonic solution below},
\begin{enumerate}
\item {For a given set of values of $E,~\lambda ~\&~ {\dot M}$, we start integration from a point asymptotically close to the horizon at $\rin=2.001$
(in units of $\rg$).}
\item {As $\rin \rightarrow 2$, $E \rightarrow ~{\cal E}=-h_{\rm in}u_t=h_{\rm in}(1-2/\rin)^{1/2}\gamma_v\gamma_\phi$.
Here $h_{\rm in}=h_{\rm in}(\thetapin,\thetaein)$ is the specific enthalpy at $\rin$.}
\item {We supply $\thetapin$.}
\item {We also supply a $\thetaein$. Then, $\vin$ is obtained as a function of the flow parameters and can be expressed as,}
$$
\vin=\left[1-\frac{(1-2/\rin)}{{\cal E}^2}\frac{\rin^3}{\{\rin^3-\lambda^2(\rin-2)\}}h_{\rm in}^2\right]^{1/2}
$$
\item {Using the values of $\thetapin,~\thetaein$ and $\vin$ we integrate Eqs.~\ref{eq:dtpdr}-\ref{eq:dvdr}, from $r=\rin$ to
outwards. As we integrate, we simultaneously check the sonic point conditions (Eqs. \ref{eq:spc1} and \ref{eq:spc2}).}
\item {If sonic point is not found, we supply another value of $\thetaein$ and repeat steps 4 and 5,
until the sonic point conditions are satisfied}.
\item {Once a sonic point is found, we integrate the equations of motion from sonic point to larger distances and obtain the
full, global, transonic two-temperature accretion solution.}
\item {After we locate one sonic point, we change $\thetaein$ again and repeat steps 4-7, in order to
check if any other sonic point exists. If found we obtain its corresponding transonic solution.}
\end{enumerate}

We note that if our supplied guess value of $\thetapin$ is unphysical, then even by
iterating with all possible values of  $\thetaein$, sonic point conditions can never be satisfied.
This is basically the modified version of the methodology adopted by \citet{lb05},
{who obtained} accretion solutions for a dissipative flow
in the single temperature regime. 

{We illustrate the procedure to find transonic solutions, enlisted above, in Figs. \ref{fig:1}a-c.
All three panels in this figure plots  Mach number ($M=v/a$) vs log $r$.
The accretion disc parameters are  $\lambda=2.5$, $E=1.000045$ and $\dot{M}=0.001\medd$
around a {BH of $10\msol$}. We would like to point out, that all these accretion disc parameters and the BH mass chosen
are for representative purpose only. These have been varied and their effect on the solution have been studied later.

{In Fig. \ref{fig:1}a, we present the method to obtain inner sonic point or $\rci$ and the transonic solution
through it. Following step 1, 2 and 3, we supply $\thetapin=7.162\times 10^{-2}$ for the aforementioned values of
$E,~\lambda$ and ${\dot M}$.
Following step 4, we start by supplying a high value of $\thetaein$ and obtain $\vin$. Then we integrate the equations
of motion (step 5). For higher values of $\thetaein$, we obtain multivalued branch (MB) of solutions.
We plot one such MB solution (green dashed-dot) corresponding to $\thetaein=11.948$.
Clearly, MB solutions are not correct. We reduce the value of $\thetaein$ (step 6) and repeat the whole procedure
up to step 5. We observe that as we reduce $\thetaein$, the MB solutions will approach the transonic solution, i.e., will shift rightward,
but in all probability we would over shoot the transonic solution
and end up with a purely supersonic branch (SB) solution (\ie when $v>a$ or $M>1$ at all $r$).
We plot a representative case of a purely SB solution (blue dashed-dot) corresponding to $\thetaein=7.731$.
When the solutions corresponding to various $\thetaein$ suddenly shifts from a MB solution to a
SB, then we know that the $\thetaein$ corresponding to a transonic solution (TS),
lies in between these two values of $\thetaein$.
We iterate on the electron temperature at $\rin$ within the range $7.731<\thetaein<11.948$ and obtain the
transonic solution (TS) (red dashed) and the sonic point is at $\rci=5.186$ (black circle). Then, by following
step 7 we obtain the complete transonic solution from $\rin$ to a large distance through $\rci$.
In Fig.~\ref{fig:1}b, we present the procedure to find the existence of the outer sonic point
for the same set of flow parameters. Following step 8 we reduce $\thetaein$ by a comparatively large value,
such that we start obtaining MB type
solutions similar to the ones we obtained while trying to locate $\rci$.
In this panel we present an example of MB solution (green dashed-dot) for $\thetaein=6.571$.
We repeat steps 4-6 and check when the solution jumps from MB to SB. This time it corresponds to $\thetaein=7.106$ (blue dashed-dot). We iterate on $\thetaein$ between the two limits  $6.571<\thetaein<7.106$, until we obtain a transonic solution 
through $\rco=3644.9$ (black star).
No other sonic point exists
for these flow parameters ($E,~\lambda,~{\dot M}$). In Fig. \ref{fig:1}c, we plot the complete set of transonic solutions,
accretion (red dashed and red solid) as well as equatorial wind solutions (red dotted)
for the disc parameters $\lambda=2.5$, $E=1.000045$ and $\dot{M}=0.001\medd$
around a BH of $10 \msol$.
The global accretion solution connecting infinity to the horizon is represented by red solid line while red dashed line represents accretion solution which is not global. We should remember that}, not all set of disc parameters produce multiple sonic points and this point will be {discussed in details in the
later sections.

\subsection{Presence of degeneracy in two-temperature transonic solutions : Method to  remove it and obtain unique transonic
solutions, invoking the second law of thermodynamics}
\label{sec:removdeg}
\begin{figure}[h]
\begin{center}
 \includegraphics[width=14.cm,height=12cm,trim={2cm 4.5cm 0.2cm 2.cm},clip]{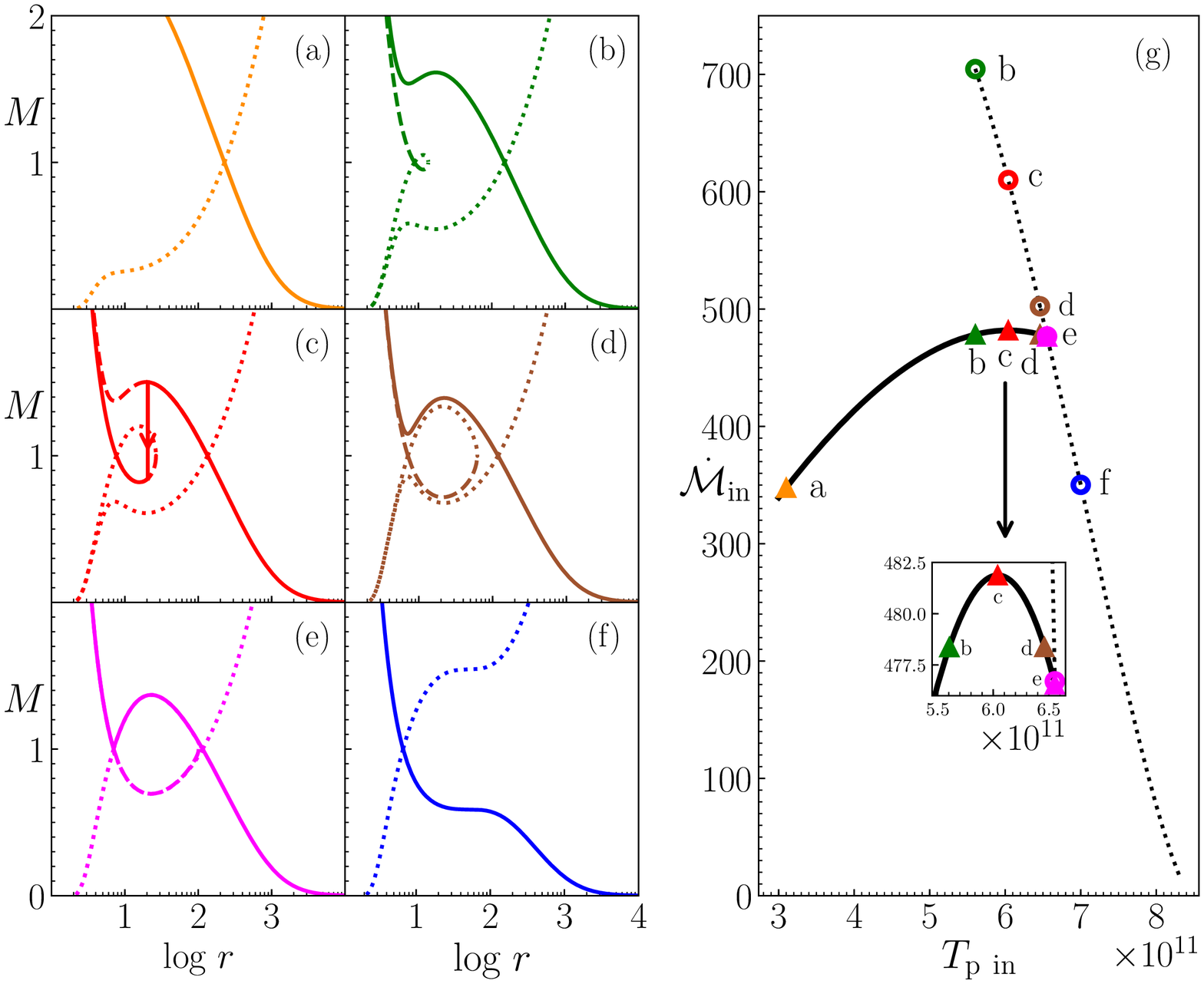}
 \caption{Left: $M$ vs log $r$ plot for various values of $\tpi$. (a) $\tpi=3.1\times10^{11} K$, (b) $\tpi=5.605\times 10^{11}K$,
 (c) $\tpi=6.04\times 10^{11}K$,
 (d) $\tpi=6.460\times 10^{11} K$, (e) $\tpi=6.554\times 10^{11} K$ and (f) $\tpi=7.0\times 10^{11} K$.
 Global solutions are represented by
 solid lines. In panel (g), $\mdotin$ vs $\tpi$ is plotted.  Solid black curve is for the solutions passing through outer sonic point,
 while dotted black curve is for solutions passing through
 inner sonic point. Panels (a)-(f) are the solutions corresponding to the points marked in right panel (g). The disc flow {parameters}
 are $E=1.0015$, $\lambda=2.6$ {and} $\dot{M}=0.02\medd$. The space time is described by a BH of {mass} $10\msol$. 
}
\vskip -0.75cm
\label{fig:2}
 \end{center}
\end{figure}

In the last section, we {laid down the procedure} to obtain transonic solution, by supplying a guess value of $\thetapin$ and
iterating with various values of $\thetaein$, until we get the sonic point.
This has been elaborately discussed in steps 1-8 of Sect.~\ref{sec:sonicptprocedure}.
{Now, if we choose a different value of} $\thetapin$ and {again follow the steps 1-8 of
Sect. \ref{sec:sonicptprocedure},} for the same {set of disc parameters ($E$, $\lambda$ \& ${\dot M}$)},
we will obtain a
different transonic solution with distinctly different sonic point properties. This means, for a given set of constants of motion,
two-temperature EoMs admit multiple transonic
solutions. Since the number of EoMs (accretion rate, momentum and energy equation) are less than the number of flow variables
(density, velocity components and two temperatures), therefore even the transonic solutions become degenerate.
From {the} second law of thermodynamics it is clear that out of all the possible {solutions}, only the solution with
{the} highest entropy {($\cal{\dot{M}}$)} should be favoured by nature \citep{sc19}.
In dissipative systems entropy is not conserved, so we measure entropy  in the region near the event horizon 
(see, Eq. \ref{eq:ear} of Sect. \ref{sec:entropy}).
For each transonic solution corresponding to a given $\thetapin$ 
we compute $\mdotin$.
{If the computed $\mdotin$ is plotted with respect to $\thetapin$ then there is a clear maxima in $\mdotin$}.
Following the second law of thermodynamics,
the solution (corresponding to a $\thetapin$) with maximum entropy {($\mdotin$)
is the physically} plausible solution. Hence, we are able to constrain the degeneracy and obtain a unique transonic
two-temperature solution for a given set of constants of
motion. 

In Figs. \ref{fig:2}a-g, we illustrate the methodology to obtain unique two-temperature transonic solution.
We choose the accretion disc flow parameters : $E=1.0015$, $\lambda=2.6$, $\dot{M}=0.02\medd$ and $\mbh=10\msol$.
As {described above}, we supply a $\thetapin$ or equivalently $\tpi$ and then we iterate on $\thetaein$ (or $\tei$)
to obtain a transonic solution.
The entropy accretion rate $\mdotin$ corresponding to the transonic solution for the particular $\tpi$ is
plotted in Fig.~\ref{fig:2}g.
The black solid line is for those $\tpi$ whose solution passes through outer sonic point ($\rco$) and black dotted line is for those passing through
inner sonic point ($\rci$). {From this figure, we select few $\tpi$s (points marked `a'--`f') which are (a) $3.1\times10^{11}$K,
(b) $5.605\times 10^{11}$K,
(c) $6.04\times 10^{11}$K, (d) $6.460\times 10^{11}$K,
(e) $6.554\times 10^{11}$K and (f) $7.0\times 10^{11}$K and plot their solutions in terms of $M$ vs log $r$, presented in Figs.~\ref{fig:2}a--f respectively. }
The global accretion solutions are represented by solid curves, dotted are wind types, while the dashed curves are
accretion solutions which are not global.
There is a range of $\tpi$ where both inner
and outer sonic points are present and is the multiple sonic point regime. For example, `b', `c' and `d' has both inner
{(marked circle)}
and outer sonic points {(marked triangle)}. For point `e', {the solutions passing through inner and outer sonic points
(magenta triangle and magenta circle), have
almost the same value of $\mdotin$ (see inset)}. Corresponding
solution is plotted in Fig.~\ref{fig:2}e, which shows that the global solution (connecting
horizon to large distances) passes through $\rco$. Now, corresponding to point `a' (orange triangle, in panel g),
the solution passes only through $\rco$ (Fig. \ref{fig:2}a),
while for point `f' (blue circle), the solution passes through $\rci$ (Fig. \ref{fig:2}f). However,
the entropy is exactly same for both these points. This means, the solution character can be completely different even if $\mdotin$ has the same value}. 
{Similarly another pair `b' and `d' also has the same entropy but different $\tpi$. Also, their corresponding solutions are significantly different, although both solutions lie in the multiple
sonic point regime}. This shows that, not only all solutions presented in the figure
have same $E,~{\dot M},~ \lambda$, but may have even same $\mdotin$ and yet the solutions
are completely different. 
In order to drive home the point even further, we present $\tpi,~\mdotin,~\rc,~v_{\rm c}, ~\&
~L\mbox{ (luminosity})$ of all degenerate solutions in Table \ref{table:1}. Some
solutions can be about four times more luminous than other solutions. It is evident from figure and table, that degeneracy in
two-temperature model is a serious problem. Observational
parameter like $L$ is quite different for {different degenerate solutions}. Any {random} choice {from the pool
of degenerate solutions} would provide us
with a completely wrong information of the system and also a wrong spectrum. Removal of degeneracy is hence important. 
Although, all the solutions presented {above} have the same energy, angular momentum and accretion rate, only one of them possess the highest
entropy (highest $\mdotin$). In this particular case, the highest entropy solution is the
one corresponding to point `c' (red triangle) in $\mdotin$-$\tpi$ curve  (Fig.~\ref{fig:2}g), and the correct, unique two-temperature accretion solution is represented
in Fig.~\ref{fig:2}c (red solid). 

\begin{table}[h!!!]
\caption{Various flow properties of the solutions plotted in Figs. \ref{fig:2}a-f. The disc parameters used are $E=1.0015$,
$\lambda=2.6$, $\dot{M}=0.02\medd$ around $\mbh=10\msol$.}
\label{table:1}
\centering
\begin{tabular}{c c c c c c c c c c c c c c }
\hline\hline
&$\tpi$ &  \multicolumn{2}{c}{$\mdotin$} &\multicolumn{2}{c}{$r_{\rm c}$ (sonic point)} & \multicolumn{2}{c}{$v_{\rm c}$
($v$ at $\rc$)} &
$L$\\
&$(\times 10^{11}K)$ & Inner & Outer & Inner & Outer&  Inner & Outer &  ($\times 10^{33}$ ergs s$^{-1}$)\\
\hline
a&3.100 & --            &350.132 &--          & 227.514 &--&  0.041 &  1.092  
\\
b&5.605 & 704.505&478.373&   8.826   &149.380 & 0.177 & 0.051  & 1.763  

\\
c&6.040 & 609.971&481.873&    7.651 & 134.135 &  0.191  & 0.054&  4.467   
\\
d&6.460 & 502.357&478.377& 7.118 &118.565  &  0.199  &  0.057&   2.457 
\\
e&6.554 &476.702&476.552& 7.027 & 11.946& 0.200 &0.058 & 2.722 
\\
f&7.000 &350.168&--&  6.672  &--& 0.207 &--&   3.020   
\\

\hline
\end{tabular}
\end{table}
\subsection{{Stability of highest entropy transonic solutions}}
\label{sec:stable}
\begin{figure}[h!!!!!!]
\begin{center}
 \includegraphics[width=7.5cm,height=7.5cm,trim={0.4cm 7.5cm 8.2cm 1.cm},clip]{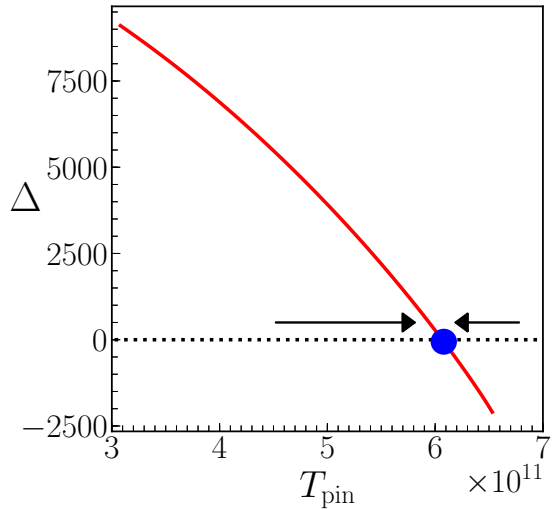}
  \caption{{Stability analysis of the unique transonic two-temperature solution with maximum entropy. The flow parameters
  used are same as Fig.~\ref{fig:2}. $\Delta=\left({d \mdotin}/{d\tpi}\right)$ is plotted against variation of $\tpi$.
  The arrows indicate that $\Delta$ converge at $\tpi=\tpim$ (blue dot) and is the stable equilibrium solution. This $\tpi$ is the
  solution with maximum entropy marked `c' in Fig.~\ref{fig:2}g.}}
\label{fig:3}
 \end{center}
\end{figure}
{In this section, we investigate the stability of the unique transonic two-temperature solution selected from the
available set of degenerate solutions. We note that the proposed unique solution is
of highest entropy and by second law of thermodynamics, nature should prefer it. Therefore, the solution should be stable.
However, because there
is a degeneracy of solutions, we need to study the stability of the proposed unique solution too. We
provide a qualitative analysis for stability of the unique two-temperature solution. \\

Let us assign $\mdotinm= {\rm{max}}(\mdotin)$ and $\tpim=\tpi$ for which $\mdotin$ is maximum. Let us further
define $\delta \tpi$ which is the difference between adjacent higher and lower $\tpi$ and $\Delta =\left({d\mdotin}/{d \tpi}\right)$ as the gradient of $\mdotin$. 
Now the change in $\mdotin$ is given by,
\begin{equation}
\delta\mdotin=\left(\frac{d\mdotin}{d\tpi}\right) \delta\tpi= \Delta   \delta\tpi
\label{eq:mdotstab}
\end{equation}
It is clear that $\Delta =0$ for any extrema of $\mdotin~-~\tpi$ curve, but the solution is said to be stable if $\tpi$ moves away from
the value $\tpim$ and the system adjusts automatically, to regain its old value.  
We prefer the graphical method to investigate the stability and the technique is similar to the first derivative test for
obtaining local extrema  \citep{melo14}. \\ \\
We plot $\Delta$ vs $\tpi$ in Fig.~\ref{fig:3}. At $\tpi < \tpim$, the figure shows that
  $\Delta>0$. Then
from Eq.~\ref{eq:mdotstab}, $\delta \mdotin>0$. So the system tends to go to higher $\mdotin$,
which we denote using a rightward arrow. Similarly,
for $\tpi > \tpim$, $\Delta<0$ which implies $\delta \mdotin<0$. Since, by second law of thermodynamics, physical system
would not like to decrease its entropy therefore for $\tpi>\tpim$, the system would tend to come back to $\tpim$.
We represent this in the figure by using a leftward arrow.
In other words, entropy can
only increase ($\delta \mdotin>0$), if $\tpi \rightarrow \tpim$ from either side of $\tpim$. Hence solution corresponding to
$\tpi=\tpim$ is stable.
Therefore, in addition to the fact that $\tpim$ corresponds to a solution with maximum entropy,
we can conclude that the solution is also stable.}

\section{Results} \label{sec:result}
Two-temperature accretion solutions are parameterized by $E$, $\lambda$, $\dot{M}$. In addition, $\betad$ and $\beta$ controls
heating and cooling. Since, we quote accretion rates in terms of 
Eddington rate, therefore the information of $\mbh$ also enters the solution. In this section we will study in
details, the two-temperature accretion solutions as well as discuss their spectral properties.
We have only analysed solutions with maximum entropy, selected from the available degenerate group of
solutions.

\subsection{General two-temperature solutions :}

\begin {figure}[h]
\begin{center}
 \includegraphics[height=10cm,width=18.cm,trim={0.1cm 10.2cm 1.2cm 2.3cm},clip]{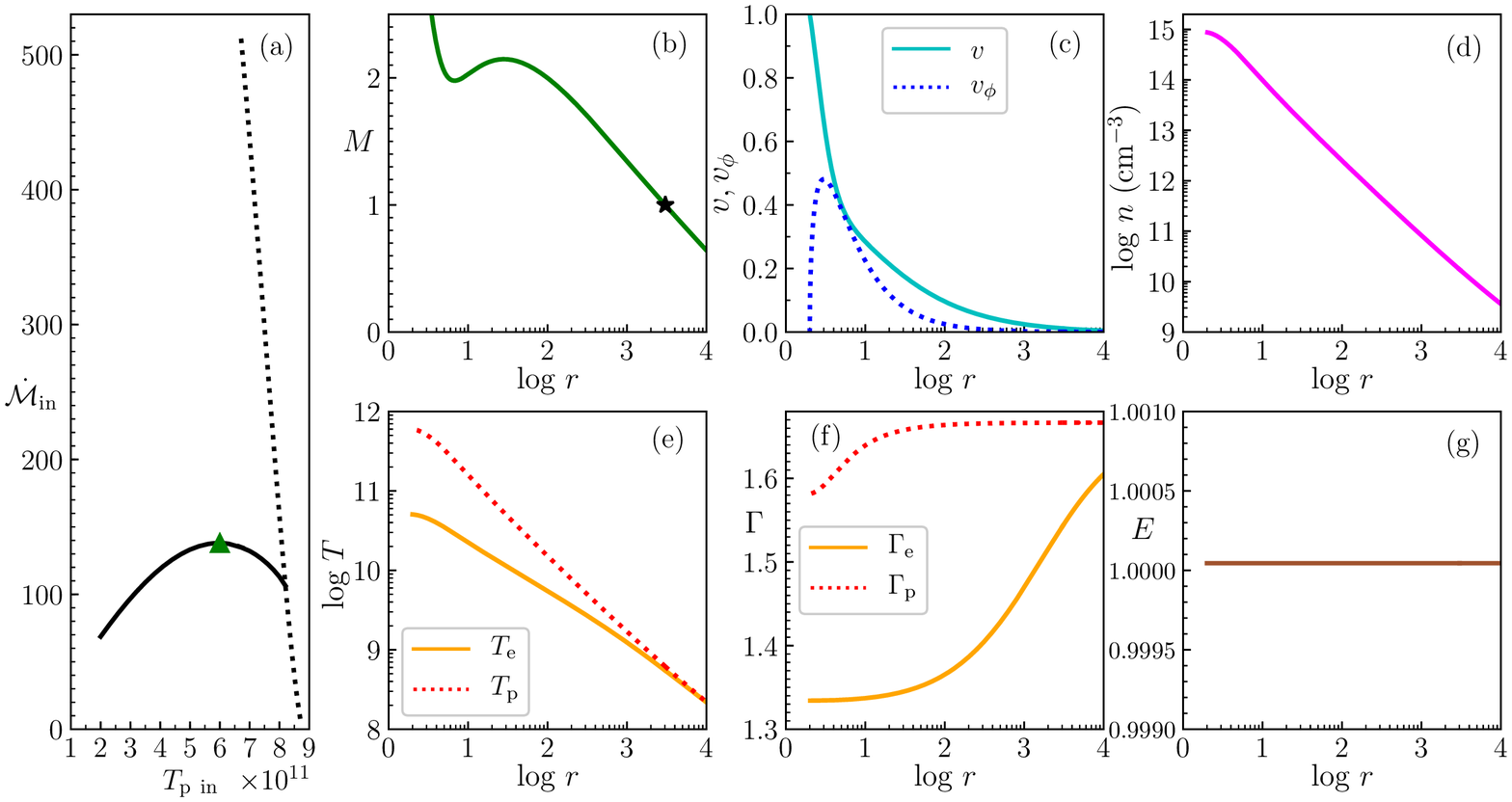}
 \caption{(a) $\mdotin$ is plotted against $\tpi$. The entropy for inner sonic point solutions (black, dotted)
and outer sonic points (black, solid) are presented. $\tpi$ marked with green triangle corresponds to maximum entropy solution.
Flow variables plotted are (b) $M$ (green, solid), (c) $v$ (cyan, solid) and $v_\phi$ (blue, dotted), (d)
log $n$ (magenta, solid), (e) $\tp$ (red, dotted), $\te$ (orange, solid), (f) $\gamp$ (red, dotted), $\game$ (orange, solid),
 (g) $E$ (brown, solid) as functions of log $r$.
  The flow parameters are $E=1.000045$, $\lambda=2.5$, ${\dot M}=0.001\medd$ and $\mbh=10\msol$. In panel b, the sonic point is
  marked with a black star.}
\label{fig:4}
 \end{center}
\end{figure}
In Fig.~\ref{fig:4}, we study a typical two-temperature transonic advective accretion disc solution. The parameters used are,
$E=1.000045$, $\lambda=2.5$,  $\dot{M}=0.001\medd$ and $M_{BH}=10\msol$. In
Fig.~\ref{fig:4}a we plot $\mdotin$ vs $\tpi$. The solution with
$\tpi=6.0 \times 10^{11} K$ has the maximum entropy (marked with green triangle) {and the corresponding $M$ vs log $r$
(green solid line)} is plotted
in Fig.~\ref{fig:4}b. The global solution passes through an outer sonic point whose position is
$r_{\rm{co}}=3040.182$ (black star). The radial three-velocity ($v$; cyan solid) in co-rotating frame and flow velocity in the
azimuthal direction ($v_\phi$; blue dotted) is plotted in Fig.~\ref{fig:4}c. Matter far away from the horizon has
negligible velocity in radial as well as in azimuthal directions. But as it approaches the BH ($r \rightarrow \rg$),
$v\rightarrow c$, thus satisfying the BH boundary condition. {On the other hand, $v_\phi$ increases with the decrease
of $r$, but maximizes at $r=3\rg$ and finally goes to zero on the horizon.}
%
This is mainly because near the horizon infall timescale is much shorter than any other timescales.
The strong gravity does not allow the matter enough time to rotate in the azimuthal direction.
In
Fig.~\ref{fig:4}d we have plotted number density (in units of cm$^{-3}$) as a function of radius. The number density increases with
the decrease in
radius, as it should be for a convergent flow. 
$\tp$ (red dotted) and $\te$ (orange solid) are plotted in
Fig.\ref{fig:4}e.
The cooling processes are dominated by electrons compared to protons.
They are however coupled by a Coulomb coupling term which
acts as an energy exchange term between the protons and electrons, as have been discussed before. This term is weak, which allows
protons and electrons to equilibrate into two different temperatures ($\tp$ and $\te$), unlike in {the} case of
single-temperature flows
where Coulomb coupling {term is assumed to be very efficient which allows}, protons and electrons
{to} attain a single temperature. Panel Fig.~\ref{fig:4}f shows that adiabatic indices of both protons
(red dotted) and electrons (orange solid), varies with the flow. This justifies our use of CR EoS.
$\gamp \sim 1.66$ and $\game \sim 1.60$ at
large distances away from the BH, hence both the species are thermally non-relativistic. When the flow approaches the BH,
$\game$ becomes 
relativistic \ie  $\game \sim 1.33$ near the horizon. We can see that $\gamp$ does not vary much but becomes
mildly-relativistic near
the horizon, owing to the higher mass of protons. In Fig.~\ref{fig:4}g, we prove that the generalized Bernoulli constant
is a constant of motion throughout the flow, even in the presence of dissipation.

\subsubsection{Emissivities and spectral properties :}

\begin {figure}[h]
\begin{center}
 \includegraphics[height=10cm,width=14cm,trim={0.cm 2.cm 1.25cm 5.1cm},clip]{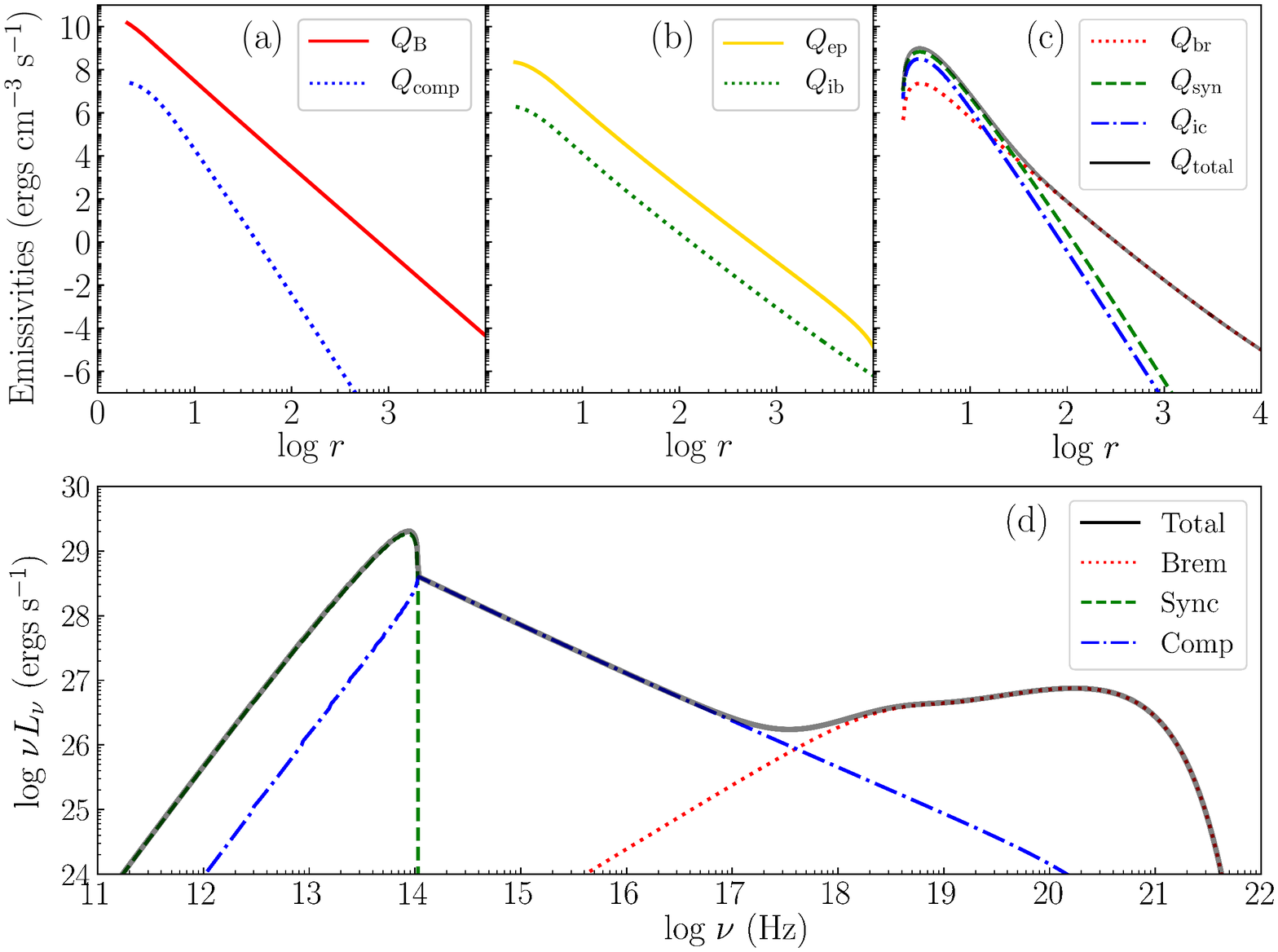}
 \caption{Top three panels shows the emissivity vs log $r$ plot for the flow presented in Fig.~\ref{fig:4}.
 Bottom panel shows
 the spectrum of the accretion flow.
}
\vskip -0.75cm
\label{fig:5}
 \end{center}
\end{figure}
In Fig.~\ref{fig:5}, we present the heating and cooling rates for the solution plotted in Fig.~\ref{fig:4}. In Fig.~\ref{fig:5}a, we plot the
heating terms. Red solid line represents the heating due to magnetic dissipation. This amount of heat is assumed to be 
equally distributed among
protons and electrons. Blue dotted line represents Compton heating of electrons. It is mainly the hard bremsstrahlung photons present 
inside the flow which leads to
heating up of the
electrons, owing to their higher energy than electrons. In Fig.~\ref{fig:5}b we plot Coulomb Coupling term (yellow solid line) and
inverse bremsstrahlung (green dotted line). 
$Q_{\rm B}$ is the strongest heating term.  

In Fig.~\ref{fig:5}c we plot the emissivities of all the cooling
processes for
electrons : bremsstrahlung (red dotted line), synchrotron (green dashed line) and inverse-Comptonization (blue dashed-dotted line). 
For the present set of disc parameters, at the outer boundary of the disc, the temperatures are non-relativistic and therefore,
bremsstrahlung emission dominates over all other processes. {In the inner regions of the accretion disc i.e., close} to the BH {horizon},
synchrotron and inverse-Comptonization becomes important and exceeds bremsstrahlung. However, inverse-Comptonization
is less than
synchrotron emission, mainly due to the low accretion rate of the flow. The total cooling of electrons is represented by a black
solid line. 
In panel Fig.~\ref{fig:5}d, we plot the spectrum for the accretion flow (black solid line). It is plotted by summing up the contributions
of all emission processes at each radius.
General and special relativistic frame transformations from fluid rest frame to the observer frame
has been taken into account while computing the spectra, including photon capture and photon bending effect due to the presence of strong
gravity. This has been elaborately discussed in Sect. (\ref{sec:spec}).
Spectrum of each emission process is also plotted. Bremsstrahlung is shown in red dotted line, synchrotron by green dashed line and
inverse-Comptonization by blue dashed-dotted line. The overall luminosity of the system is low, $L=2.536\times10^{29}$ ergs s$^{-1}$,
with an radiative efficiency of $\eta_{\rm r}=1.957 \times 10^{-5}$. It may be noted that, efficiency is defined as
$\eta_{\rm r}=L/(\dot{M}c^2)$. The spectral index is $\alpha=1.744$.




\subsection{Contributions of different regions of the accretion disc to the overall spectrum :}
\begin{figure}[h!!!!!!!]
\begin{center}
\includegraphics[width=18.0cm,height=6.5cm,trim={0.cm 16.cm 0.cm 0.cm},clip]{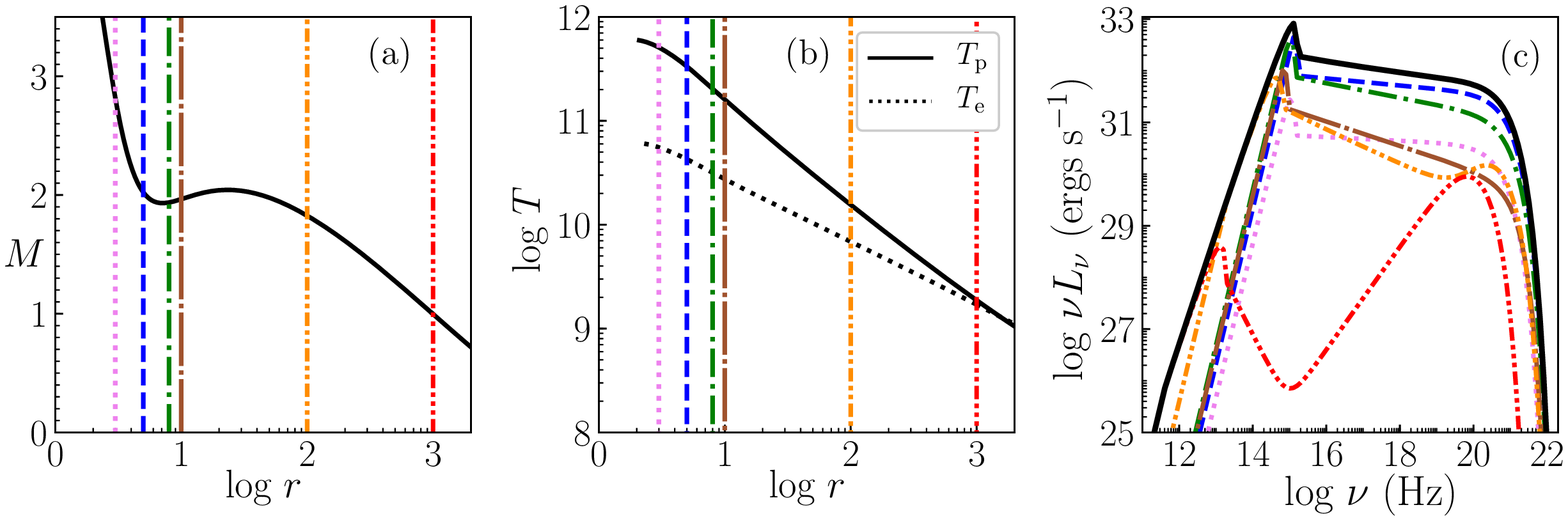}
\caption{\label{fig:6}(a) $M$ and (b) log $T$ vs log $r$ and (c) total spectrum (black solid) and contribution from 
various length scales of the accretion disc, $2-3\rg$ (magenta, dotted),
$3-5\rg$ (blue, dashed), $5-8\rg$ (green, short-dashed-dotted), $8-10\rg$ (brown, long-dashed-dotted), $10-100\rg$
(orange, dashed-double-dotted) and $100-1000\rg$ (red, dashed-triple-dotted). Flow parameters are
$E=1.0002,~\lambda=2.48,~{\dot M} = 0.05\medd$ and $\mbh = 10\msol$. }
\vskip -0.75cm
\end{center}
\end{figure}

In Fig.~\ref{fig:5}d, we showed the contribution of all the emission processes in the total broad band
continuum spectrum. {However, in the following we would like to investigate the contribution of various regions of an accretion
disc in the overall spectrum.}
We chose a set of flow parameters $E=1.0002$, $\lambda=2.48$, ${\dot M} = 0.05\medd$ and $\mbh = 10\msol$.
In Fig.~\ref{fig:6}a, we plot the Mach number $M$ of the accretion flow
and in Fig.~\ref{fig:6}b, we plot $\tp$ (black solid) and $\te$ (black dotted) as a function of log $r$. In panels (a) and (b), we
indicate various regions with vertical lines
which represents accretion  disc section from $2-3\rg$ (magenta, dotted), $3-5\rg$ (blue, dashed), $5-8\rg$ (green, short-dashed-dotted), $8-10\rg$ (brown, long-dashed-dotted), $10-100\rg$ (orange, dashed-double-dotted)
and $100-1000\rg$ (red, dashed-triple-dotted). The spectra from all these regions are separately over plotted in Fig.~\ref{fig:6}c,
the colour coding of the spectra matches the region from which they are computed. The black curve
represent the overall spectrum for the disc parameters stated above. The contribution from the region $r=10^3-10^4\rg$ is
too low in the overall spectrum and therefore is not plotted, in order to avoid cluttering the figure.
The spectrum computed from the region $2-3\rg$ is low inspite of high values of $n$ and $\te$, since significant number of photons
emitted from that region, are captured by the BH.
Most of the high energy emission is contributed by accreting matter from the region between $3-5\rg$ and $5-8\rg$, and the
low-energy end
of the spectra from this region is always around and above $10^{12}$Hz. We have tabulated these details and other spectral
properties in Table \ref{table:2}. We can conclude from the table that $\sim 90\%$ of the emission comes from a region
$< 10\rg$ of the accretion disc.
The lower energy part of the spectrum is mostly contributed by the outer part of the disc.
Since we have only considered advective disc, the spectrum is hard and the radiative efficiency for this particular {set of}
disc parameters is less than $1\%$. 

\begin{table}[h!!!]
\caption{{Spectral properties of the regions marked in Fig. \ref{fig:5}}}
\label{table:2}
\centering
\begin{tabular}{c c c c}
\hline\hline
Colour &Region (in $\rg$) &  $\%$ of $L_{\rm tot}$ &$\alpha$ \\
\hline
Magenta&2-3 &  3.597 &1.044 
\\
Blue&3-5 &  45.810 & 1.072
\\
Green&5-8 &  35.250 & 1.141
\\
Brown&8-10 & 6.965 & 1.236
\\
Orange&10-100 &  7.845 & 1.323 
\\
Red&100-1000 & 0.286 & 2.392 
\\
-- &1000-10000 &  0.247 & 2.233x10$^{-5}$
\\
\hline
\end{tabular}
\end{table}

\subsection{Dependence of accretion solutions and corresponding spectra with energy and angular momentum:}
\begin{figure}[h!!!!!!!]
\begin{center}
\includegraphics[width=10.5cm,height=10.5cm,trim={0.1cm 0.6cm 1.cm 1.cm},clip]{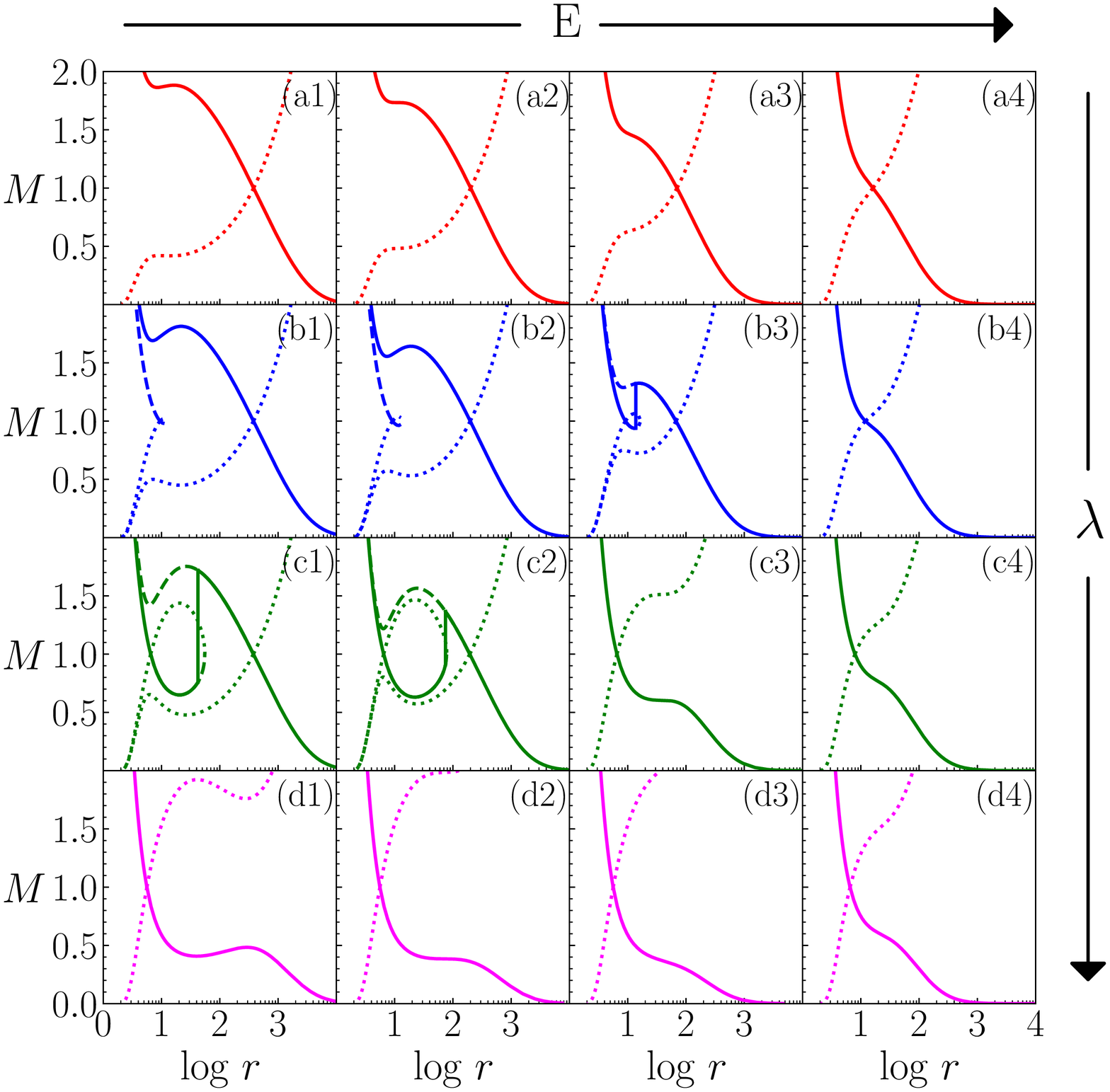}
\caption{\label{fig:7}Variation of solutions, $M$ as a function of log $r$ with variation of $E$ and $\lambda$.
From left to right specific energy increases as
$E=1.0005$, $1.001$, $1.003$ and $1.01$. From top to bottom the angular momentum increases as
$\lambda=2.40$,
$2.55$, $2.70$ and $2.85$. Other parameters are ${\dot M} = 0.01\medd$ and $\mbh = 10\msol$.}
\vskip -0.75cm
\end{center}
\end{figure}

\begin{figure}[h!!!!]
\begin{center}
\includegraphics[width=10.5cm,height=10.5cm,trim={0.cm 0.1cm 1.cm 1.cm},clip]{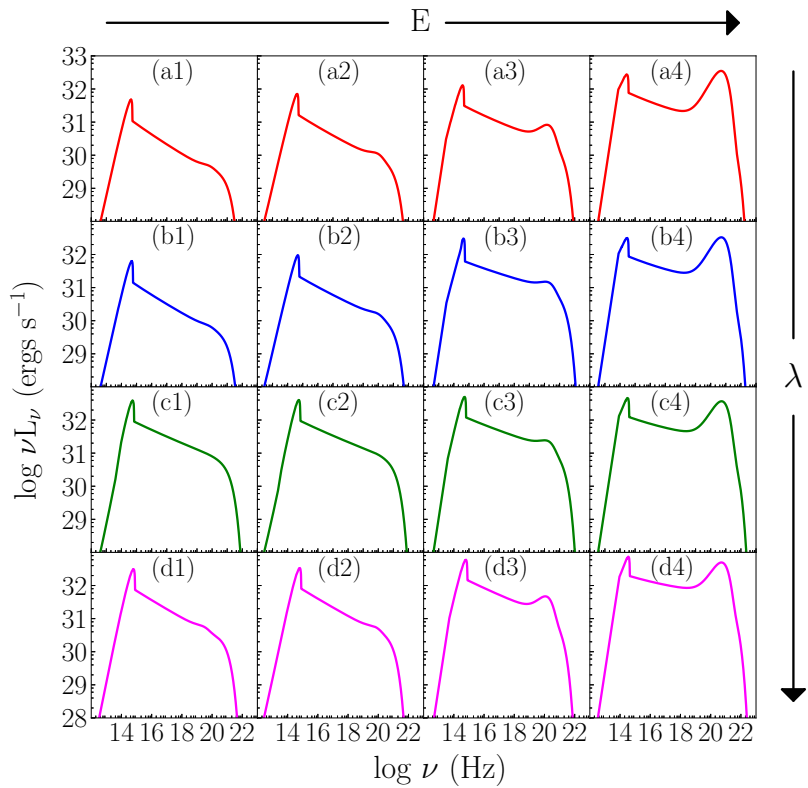}
\caption{\label{fig:8} Variation of spectrum with $E$ and $\lambda$.
The set of values for $E$ and $\lambda$ and other
parameters are same as that in Fig.~\ref{fig:7}.}
\vskip -0.95cm
\end{center}
\end{figure}

In Figs. \ref{fig:7} and \ref{fig:8}, we investigate the dependence of {accretion solutions} and the corresponding spectra
on $E$ and $\lambda$ for a $10\msol$ BH with $\dot{M}=0.01\medd$.
In the figures, $E$ increases from left to right and the values are $E=1.0005$, $1.001$,$1.003$ and $1.01$.
While, as we go from top to bottom, $\lambda$ increases as $\lambda=2.40$, $2.55$, $2.70$ and $2.85$. In short, $E$ changes along
the row while $\lambda$ changes along the column. Low angular
momentum flows ($\lambda=2.40$) behave as Bondi flow, possessing single sonic point, through which the global solution passes
(see Figs.~\ref{fig:7}a1-a4), irrespective of the value of $E$. As angular momentum increases, rotation head of the specific energy
($E$) of the flow play a significant role inside the system and multiple sonic points form in an
appreciable section of the parameter space. For $\lambda=2.55$ (Figs.~\ref{fig:7}b1-b3), multiple sonic point exists in a
large range of $E$.
In Figs.~\ref{fig:7}b1-b2, the global solution (blue, solid) passes through the outer sonic point whereas in panel Fig.~\ref{fig:7}b3,
the solution (blue, solid) harbours a shock
and passes through both inner and outer sonic points. In Fig.~\ref{fig:7}b4, only a single sonic point exist. This is mainly due to the
fact, that for flows with higher
energy, the distribution of sound speed $a$ is generally higher compared to flows with lower
$E$. Therefore, the flow can only become transonic, when $v$ increases significantly, which can happen
only very close to the BH. {For low values of $E$, the sound speed distribution $a(r)$ is comparatively low.
Therefore the sonic points form further out, because the flow becomes transonic whenever the infall velocity $v(r)$
attains moderately high values}.
Angular momentum has different effect on the flow structure.
If we increase $\lambda$, {then the distribution of $v_\phi(r)$ increases. Higher values of $v_\phi$
restricts the increase of $v$ to moderate values, except near the horizon}. So for {flows with higher} $\lambda$,
the sonic points shift
towards the BH. For even higher $\lambda \geq 2.85$, only inner sonic point exists irrespective of the value of $E$
(see Figs.~\ref{fig:7}d1-d4).
In Figs.~\ref{fig:7}c1-c2 which are for $\lambda=2.70$, shocks form even at low energies. {If one compares with single
temperature accretion discs \citep{cc11,kc14,
ck16,kc17}, it is clear that multiple sonic points form in a much smaller range of energy-angular momentum parameter space
of a two temperature accretion disc, and is shown in Figs.~\ref{fig:7}a1-d4. }

Figure \ref{fig:8}, shows the corresponding spectra which spans from $10^{12}-10^{22}$Hz. As a general trend, with the increase in
$\lambda$ of the system, luminosity increases, since matter gets enough time to radiate. But the spectral shape and slope (arising
because of inverse-Comptonization) {remains roughly the same}, except for the solutions which harbours shock.
The spectral slope is flatter
in case of shocked solutions. 
With the increase in $E$, thermal energy of 
the system increases, emission is hence higher. The spectral shape and slope 
is visibly changed. Bremsstrahlung 
emission, the broad peak in the higher frequency range, is increased with the increase in $E$ (left to right), while angular
momentum seems to have little effect on this particular radiative process. 
Since in this case we are dealing with a flow with low accretion rate, spectrum is relatively soft as inverse-Comptonization is
not important.

\subsection{Shocked solution, spectra and the parameter space:}
In Figs.~\ref{fig:7}b3,c1,c2, the accretion solutions admit stable shocks. As discussed before, for low $\lambda$, a flow admits
only one sonic point (Figs.~\ref{fig:7}a1-a4). But as 
$\lambda$ increases, the flow possess multiple sonic points. A flow can pass through both the sonic points
only when the shock conditions are satisfied (see, Sect. \ref{sec:shockcond}). With the increase in $\lambda$, the centrifugal
term increases, and the twin effect of the centrifugal and the thermal term can restrict the infalling matter, leading to a
centrifugal pressure mediated shock transition.
In the following section, we will analyse shocks present in two-temperature accretion flows.
\begin {figure}[h]
\begin{center}
 \includegraphics[height=11cm,width=12.cm,trim={1.1cm 3.58cm 8.1cm 5.5cm},clip]{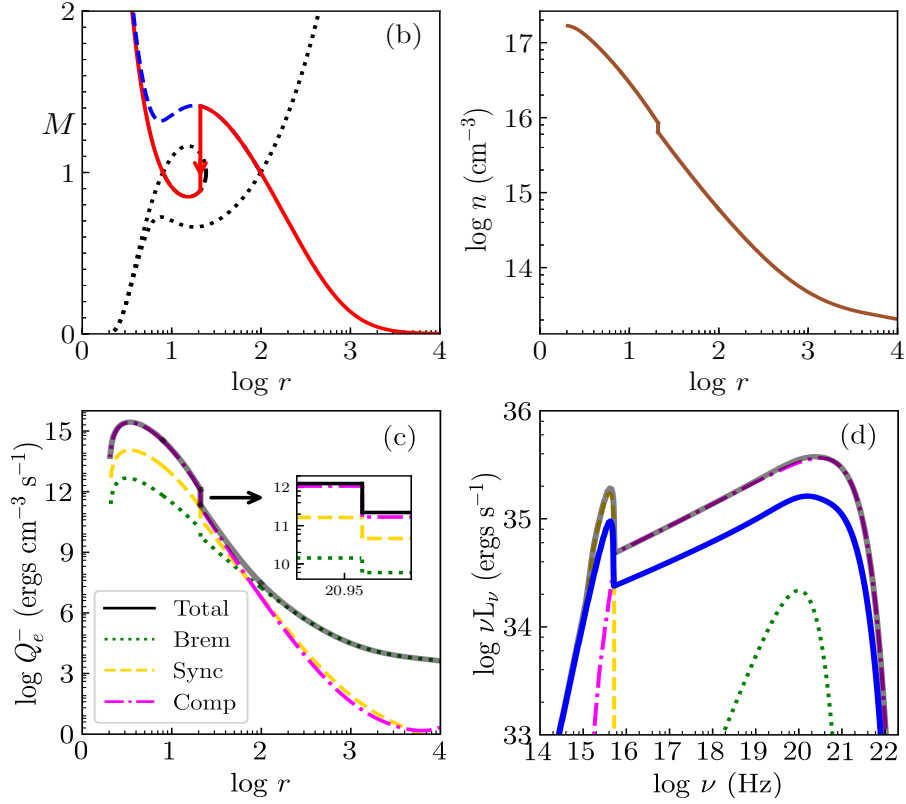}
 \caption{A typical shocked solution (a) with its corresponding number density (b), emissivities (c) and spectrum
 (d) is presented. The parameters taken are $E = 1.002$, $\lambda=2.58$, ${\dot M} = 0.2\medd$ and $\mbh = 10\msol$. 
}
\vskip -0.5cm
\label{fig:9}
 \end{center}
\end{figure}

In Fig. \ref{fig:9}a, we present a typical shock solution for the parameters $E = 1.002$, $\lambda=2.58$, ${\dot M} = 0.2\medd$ and
$\mbh = 10\msol$. For these parameters, the flow possess multiple sonic points.
Blue dashed line is for the accretion solution
passing through the outer sonic point $\rco$. When this solution becomes supersonic it encounters a shock at $\rsh =20.952$. Then it
jumps to the subsonic branch and enters the BH supersonically after crossing through the inner sonic point $\rci$. The global
solution
is represented with a red solid line. The compression ratio ($R=u^r_-/u^r_+$, $\pm$ implies post and pre-shock quantities,
respectively)
is $1.459$ in this case. In Fig.~\ref{fig:9}b, we plot the number density as a function of $r$. At the shock there is an increase in
number density of both protons and electrons equally. This leads to increased
cooling in the system which is evident from Fig.~\ref{fig:9}c, where we plot emissivities of various cooling processes related to
electrons. The corresponding spectrum is
plotted in Fig.~\ref{fig:9}d. In Figs.~\ref{fig:9}c-d, bremsstrahlung is represented using dotted green line,
synchrotron in dashed yellow
and inverse-Comptonization in dashed-dotted magenta, while the total cooling is represented by solid black line. The accretion rate of
the system is high, so the spectrum is mainly dominated by inverse-Comptonization, especially in the post-shock region.
In Fig.~\ref{fig:9}d, the total spectrum of the system is plotted in black, while super imposed on it is the spectrum (blue solid)
of the shock-free solution (blue dashed of panel a).
The luminosity of the system is $2.831\times 10^{36}{\rm~ergs~s}^{-1}$, which corresponds to an efficiency ($\eta_{\rm r}$) of $1.09\%$, while
for a shock-free branch (blue dashed), the luminosity would have been $1.293\times 10^{36}{\rm~ergs~s}^{-1}$ and $\eta_{\rm r}=0.50\%$.
Because of the shock, the luminosity and hence the efficiency of the system doubled. However,
it seems that there is no special spectral signature of shock in accretion flow, except that the luminosity of the
power-law part of the spectrum increases. This is also evident from Figs. \ref{fig:8}b3, c1 and c2.
\begin {figure}[h]
\begin{center}
 \includegraphics[height=7cm,width=7cm,trim={0.cm 0cm 1.cm 1cm},clip]{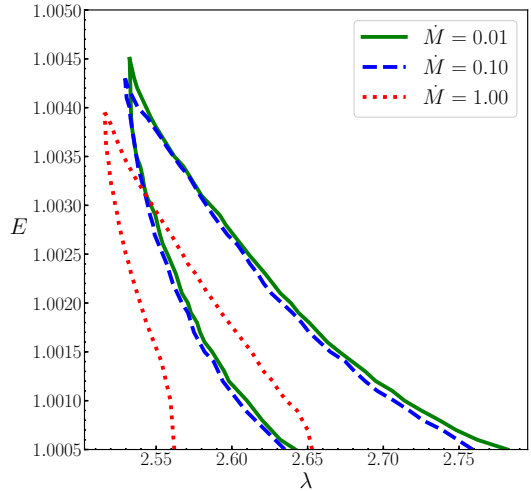}
 \caption{Shock parameter space for $\dot {M}=0.01\medd$ (green, solid), $0.10\medd$ (blue, dashed) and $1.00\medd$ (red, dotted)
 around a $10\msol$ BH. 
}
\vskip -0.5cm
\label{fig:10}
 \end{center}
\end{figure}

The bounded region in $E$-$\lambda$ space in Fig.~\ref{fig:10}, represents the shock parameter space, \ie a flow with $E$, $\lambda$
values from the bounded region for the given ${\dot M}$, will under go a stable shock transition.
Each bounded region or shock-parameter space is characterized by different
accretion rates : ${\dot M}=0.01$ (green, solid), ${\dot M}=0.1$ (blue, dashed) and ${\dot M}=1.0$ (red, dotted) around a $10\msol$ BH.
We can see that as $\dot{M}$ increases, the parameter space decreases and shifts to the lower angular momentum side.
{High value of ${\dot M}$ implies higher rate of cooling and therefore, much hotter flow at the outer boundary
can accrete and form the disc. And hence even for lower $\lambda$, the centrifugal term in conjunction with the thermal term
can resist the infall to produce an accretion shock. That is why for higher ${\dot M}$, the shock parameter space shifts to the
lower $\lambda$ values.}
For low ${\dot M}$, the shock parameter space is almost similar, it significantly changes
only in presence of high accretion rates. More interestingly, it is clear that, an accretion flow with high accretion rate may
also harbour accretion shocks.

\subsection{Dependence of spectrum on $\beta$ :}
\label{sec:betasync}
\begin {figure}[h]
\begin{center}
 \includegraphics[height=6cm,width=6cm,trim={0.7cm 5.cm 13.cm 9cm},clip]{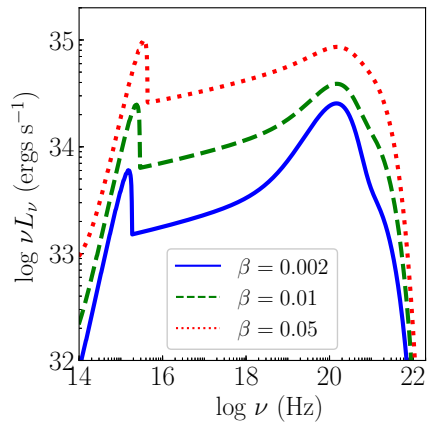}
 \caption{Change in spectra with increase in $\beta=0.002$ (blue, solid), $0.01$ (green, dashed) and $0.02$ (red, dotted).
 Other parameters used are $E=1.003$, $\lambda=2.54$ and $\dot{M}=0.1\medd$ in an accretion disc around $\mbh=10\msol$.
}
\vskip -0.5cm
\label{fig:11}
 \end{center}
\end{figure}
$\beta$ controls the magnitude of stochastic magnetic field inside the flow. Any change in it would lead to the change in synchrotron
emission from electrons and eventually change the radiation due to inverse-Comptonization. Hence, the spectra that an observer would see,
significantly depends on the value of $\beta$. In Fig.~\ref{fig:11}, we plot the change in spectra with change in $\beta$ for the flow
parameters $E=1.003$, $\lambda=2.54$, $\dot{M}=0.1\medd$ and  $\mbh=10\msol$. We have varied $\beta$: $0.002$ (blue, solid), $0.01$
(green, dashed) and $0.02$ (red, dotted). The present flow has high accretion rate where cooling is more pronounced. Even for low
$\beta$, the power law signature in the spectrum arising due to inverse-Comptonization, is hard. However, dominant emission comes
from bremsstrahlung as can be inferred from its bump at higher frequency regime. As we increase $\beta$, the bump feature vanishes.
This is mainly due to the fact that with the increase in $\beta$, synchrotron emission and hence inverse-Comptonization increases
more as compared to bremsstrahlung which is independent of the magnitude of the magnetic field in the flow. The synchrotron
turnover frequency also shifts to higher frequencies with the increase in $\beta$.
\subsection{Dependence of solutions and spectra on $\beta_{\rm d}$ :}
\label{sec:betad}
\begin {figure}[h]
\begin{center}
\includegraphics[height=12cm,width=13cm,trim={0.cm 3.cm 1.cm 0cm},clip]{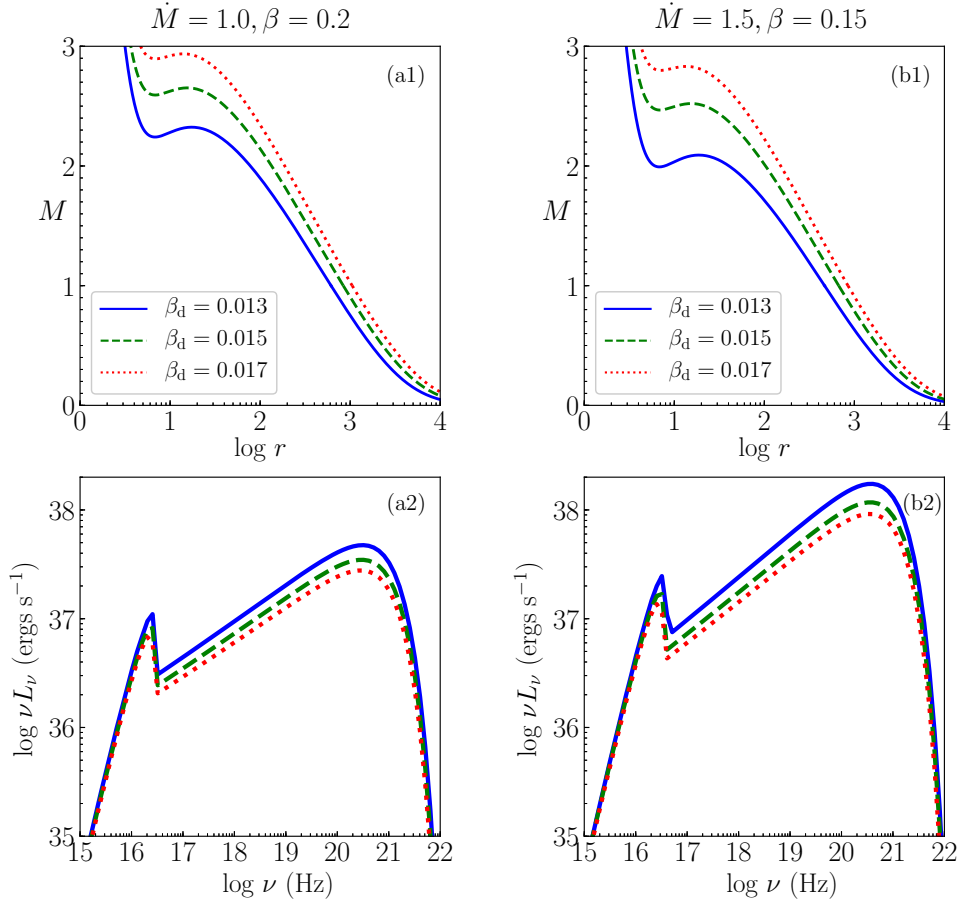}
\caption{Plotted are the accretion solutions (a1, b1) and their corresponding spectra (a2, b2) for a flow with $E=1.001$,
$\lambda=2.61$ around $\mbh=10\msol$.
Various curves are for $\betad=0.013$ (blue, solid), $\betad=0.015$ (green, dashed) and $\betad=0.017$ (red, dotted).
The accretion rates and ratio of magnetic to gas pressure are chosen are ${\dot M}=1.0\medd$, $\beta=0.2$ (a1, a2) and
${\dot M}=1.5\medd$, $\beta=0.15$ (b1, b2).
}
\vskip -0.5cm
\label{fig:12}
 \end{center}
\end{figure}

Figure \ref{fig:5}a showed that, magnetic dissipation is a more efficient heating mechanism compared to Compton heating as well as
Coulomb heating of electrons.
In Figs. \ref{fig:12}a1-b2, we vary
$\betad$ which controls magnetic dissipation. We compared the solutions and resultant spectra for $\betad=0.013$ (blue, solid),
$0.015$ (green, dashed) and $0.017$ (red, dotted). For Figs.~\ref{fig:12}a1,a2, we chose higher ratio between magnetic and
gas pressure, $\beta=0.2$ and accretion rate ${\dot M}=1.0\medd$. For Figs.~\ref{fig:12}b1,b2,
we select ${\dot M}=1.5\medd$ and $\beta=0.15$. For both the cases, we have $E=1.001$, $\lambda=2.61$ and $\mbh=10\msol$.
The sonic point, luminosities and spectral index of the accretion flows are given in Table \ref{table:3}. 
Evidently, luminosity and hence efficiency, decreases with increasing $\betad$ but increases with increasing ${\dot M}$.
If the dissipative heating is higher (i.e., higher $\betad$), then matter with lower temperature at large $r$
may achieve the same $E$. Therefore,the  accretion flow would have an overall lower temperature and would emit less.
That is exactly, what is observed in Figs.~\ref{fig:12}a2,b2, where the luminosity goes down with the increase in $\betad$.
The spectrum is decisively hard for super-Eddington accretion rates. It means that no single flow parameter
can dictate whether the spectrum will be hard or soft, instead all the flow parameters together contribute for the final outcome.
However, it is clear that $\betad$ do influence the emitted spectra and luminosity significantly. 

\begin{table}[h!!!]
\caption{Various flow properties of the solutions plotted in Fig. \ref{fig:12}. }
\label{table:3}
\centering
\begin{tabular}{c c c c c}
\hline\hline
Parameters & $\betad$ &$\rc$&$L$ & $\alpha$
\\
& & & $\times 10^{38}$ (ergs s$^{-1}$) &
\\
\hline
{$\dot{M}=1.0\medd$ and $\beta=0.20$} & 0.013 & 613.365 & 2.523 & 0.672
\\ & 0.015 & 838.022& 1.877 & 0.680
\\ & 0.017 & 1053.310& 1.514 & 0.683
\\
\hline
{$\dot{M}=1.5\medd$ and $\beta=0.15$} & 0.013 &466.718 &  8.31 & 0.606
\\ & 0.015 & 669.746 &5.726 & 0.619
\\ & 0.017 & 849.084 & 4.518 & 0.625
\\
\hline
\end{tabular}
\end{table}
\subsection{{Possibility of pair production and pion production}}
{By now we have investigated how various factors can affect the two-temperature solutions and resulting spectra. However,
it may be noted that, we have ignored pair production from particle-particle interactions in the accretion disc
or from accretion disc radiations. We have also ignored the production of gamma-rays due to high energy interactions
like pion decay. We assumed that these processes will not significantly affect the solutions. In Appendix
\ref{app:pair}, we investigate the pair production processes a posteriori. We compare the number densities of protons
$\np$ with positrons $n_{e^+}$ (Figs.~\ref{fig:b1}a1, b1) generated through photon interactions produced in accretion discs as well as compare the total emissivity $Q_{\rm total}$ with pair annihilation emissivity $Q_{\rm ann}$ (Figs.~\ref{fig:b1}a2, b2).
We consider two sets of accretion disc parameters (1) ${\dot M}=1.0,~ \beta=0.2$ (Figs.~\ref{fig:b1}a1, a2)
and (2) ${\dot M}=1.5,~ \beta=0.15$ (Figs.~\ref{fig:b1}b1, b2). Rest of the parameters common in both the cases
are $\betad = 0.013$, $E=1.001$ and $\lambda=2.61$. These two accretion disc cases are described around a BH of $10\msol$.
After the posteriori calculations, elaborately discussed in Appendix \ref{app:pair}, we can conclude that,
$n_{e^+} \ll \np$ and $Q_{\rm ann} \ll Q_{\rm total}$. \\
In Appendix \ref{app:pion} we compute the production of pions ($\pi^0$) a posteriori and the gamma rays emitted due to its decay.
We plot log $\tp$ vs log $r$ in
Figs.~\ref{fig:c1}a1, b1 and the corresponding spectra in Figs.~\ref{fig:c1}a2, b2. We study the generation
of pions and gamma ray photons for two cases (1) by varying accretion rate (${\dot M}=0.01$ : red, dotted, $0.1$ : green, dashed
and $1.0$ : blue, solid) around a BH of $\mbh=10\msol$ (Figs.~\ref{fig:c1}a1, b1) and (2)  by varying mass of the BH
($\mbh=10^2$ : blue, solid, $10^4$ : green, dashed and $10^6$ : red, dotted), keeping accretion rate ${\dot M}=0.1$ constant.
The other disc parameters are $E=1.0007,~\lambda=2.61,~\beta=0.01,~\&~\betad=0.001$. Luminosity for higher ${\dot M}$ is higher
and so is the
gamma-ray produced by decay of pions. Same trend is observed when we increase the BH mass. However, the gamma ray luminosity is
always  $< 10^{-5}$ times that of the total luminosity. Elaborate discussion on these two cases have been made in
Appendix \ref{app:pion}. We can conclude that consideration of pair production or pion decay will not affect the accretion
solutions and the spectra, significantly.
}

\subsection{Dependence on $\dot{M}$ and $\mbh$ :}
\begin {figure}[h]
\begin{center}
 \includegraphics[height=6cm,width=13cm,trim={0.cm 10.4cm 1.cm 4.1cm},clip]{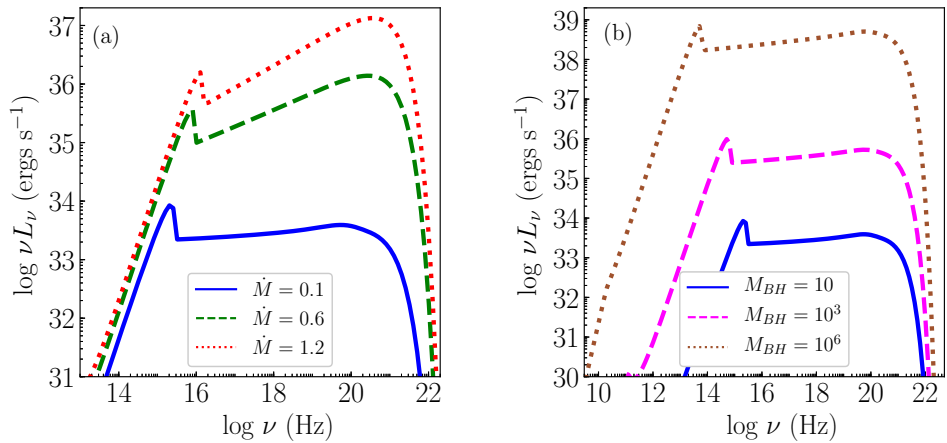}
 \caption{Spectra from (a) $\mbh=10 \msol$ for different accretion rates ${\dot M}=0.1\medd$ (blue, solid),
 ${\dot M}=0.6\medd$ (green, dashed) and ${\dot M}=1.2\medd$ (red, dotted); (b) ${\dot M}=0.1\medd$ but around
 $\mbh=10\msol$ (blue, solid), $\mbh=10^3\msol$ (magenta, dashed) and $\mbh=10^6\msol$ (brown, dotted).
 Other disc parameters are $E=1.001$ and $\lambda=2.4$.
}
\vskip -0.5cm
\label{fig:13}
 \end{center}
\end{figure}

In Fig.~\ref{fig:13}a, we plot the continuum spectra for ${\dot M}=0.1\medd$ (blue, solid),
 ${\dot M}=0.6\medd$ (green, dashed) and ${\dot M}=1.2\medd$ (red, dotted) from a disc around a BH of $10\msol$.
 The disc becomes brighter as ${\dot M}$ increases, even the efficiency also increases. The spectra also
 becomes harder, mainly because the inverse-Compton output increases with the increase in
 number density of hot electrons inside the flow. However, the range of frequency $\nu$ on which the spectrum is distributed
 do not
 increase appreciably with the increase in accretion rate. Corresponding spectral properties are presented in Table~\ref{table:4}:
 
 \begin{table}[h!!!]
\caption{Various properties of the spectra plotted in Fig.~\ref{fig:13}a. }
\label{table:4}
\centering
\begin{tabular}{c c c c c c c c c c c c c c c c}
\hline\hline
$\dot{M}$&$L$ & $\eta_{\rm r}$&$\alpha$
\\
($\medd$)&(ergs s$^{-1}$) &($\%$)
\\
\hline
0.1&  4.731 $\times 10^{34} $ &  0.037 & 0.939
\\
0.6& 8.301 $\times 10^{36} $ &1.068 & 0.720
\\  
1.2& 6.684$\times 10^{37}$&4.298 & 0.627
\\
\hline
\end{tabular}
\end{table}

In Fig. \ref{fig:13}b, we plot spectra from discs
 with the same accretion rate ${\dot M}=0.1\medd$, but around different $\mbh$ which are $10\msol$ (blue, solid), $10^3\msol$ (magenta, dashed) and $10^6\msol$ (brown, dotted).
 The more massive the black hole, the
 disc is brighter since absolute accretion rate increases. In addition the spectrum spans over a larger range of $\nu$,
 with significant emission from radio to $\gamma$ rays. It may be noted, higher $\mbh$ results in a more broadband spectra,
 the disc becomes more luminous but the spectral index do not change much. The spectral properties are presented in
 Table \ref{table:5}:
 \begin{table}[h!!!]
\caption{Various properties of the spectra plotted in Fig. \ref{fig:13}b.}
\label{table:5}
\centering
\begin{tabular}{c c c c c c c c c c c c c c c c}
\hline\hline
$M_{BH}$&$L$ & $\eta_{\rm r}$&$\alpha$
\\
($\msol$) & (ergs s$^{-1}$) &($\%$)
\\
\hline
10&  4.731 $\times 10^{34} $ &  0.037 &0.939
\\ 
$10^3$&6.426 $\times 10^{36} $  &   0.049 &0.936
\\   
$10^6$&6.106$\times 10^{39}$&  0.047 & 0.915
\\
\hline
\end{tabular}
\end{table}

\subsubsection{Luminosity, efficiency and spectral index of two-temperature flows :}
\begin{figure}[h]
\begin{center}
\includegraphics[width=14cm,height=7.5cm,trim={0.0cm 3.5cm 0.0cm 9.cm},clip]{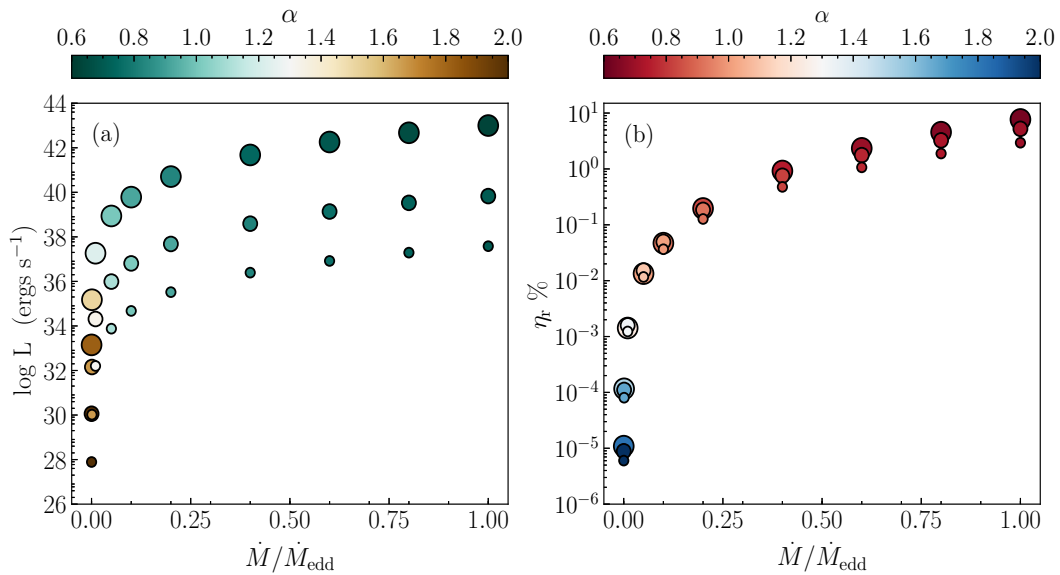}
\caption{\label{fig:14} (a) Variation of bolometric luminosity (in ergs s$^{-1}$) and (b) efficiency (in $\%$) as a function of
$\dot{M}$ (in units of Eddington rate, $\dot{M}_{\rm{edd}}$). Color bar indicates the spectral index ($\alpha$).
BHs of different masses : $10\msol$ (small circle), $10^3\msol$ (medium circle) and $10^6\msol$ (largest circle) are represented with increasing sizes of the circles.
The parameters used here are $E=1.001$ and $\lambda=2.4$.}
\end{center}
\end{figure}

\noindent In Fig.~\ref{fig:14}a, we have calculated the luminosities and in Fig.~\ref{fig:14}b we plotted the efficiency
of the accretion of matter onto BHs of different masses
($10\msol$, $10^3 \msol$ and $10^6 \msol$) as a function of accretion rate ($\dot{M}$). Size of the circles are in order of
increasing
value of BH mass. Parameters used are $E=1.001$ and $\lambda=2.4$. 
It may be noted that luminosity
rises steeply with the increase in accretion rate of the system, for all BH masses. More the supply of matter, more would
be the conversion
of it into energy. However, at higher accretions rates, luminosities approach asymptotic values.  
Radiation emitted by accretion disc, is the effect of conversion of gravitational energy released in the act of accretion,
into electro-magnetic radiation.
So as ${\dot M}$ increases, emission increases due to increased supply of matter. However, it cannot emit
more than the energy obtained from the accretion process and therefore, it reaches a ceiling around $\eta \sim 10\%$.
But it is apparent from Fig. \ref{fig:14}b, that efficiency 
is slightly affected by the BH mass. The spectral index ($\alpha$) is represented as color bar over both Figs. \ref{fig:14}a,b. It changes visibly with the increase in accretion rate of the system (also, see Fig. \ref{fig:13}a and Table \ref{table:4}) but do not change much with the change in BH mass (also, see Fig. \ref{fig:13}b and Table \ref{table:5}).

\section{Discussion and Conclusions}
\label{sec4}

In this paper, we studied {solutions of two temperature} accretion discs around {non-rotating} BH. 
{It may be noted that, the spin of the BH may play an important role in jet generation via a process called
Blandford-Znajek mechanism \citep{bz77}, however, accretion is still the primary mechanism to explain the observed
luminosities. And a proper two temperature accretion solution is the best way to obtain the spectra from such systems}.
Two temperature
equations produce degenerate set of solutions even when they have the same set of disc parameters like generalized Bernoulli
parameter
($E$), accretion rate ($\dot{M}$ and angular momentum ($\lambda$).
For the given set of disc parameters, a choice of proton temperature may
produce a transonic solution through outer sonic point, some other choice of the temperature will produce a solution
through inner sonic point, while some other will produce solutions which undergoes shock transition.
The resulting radiation also vary accordingly. Infact some solutions may
be four times more luminous compared to some other solutions (see Fig.~\ref{fig:2} and Table \ref{table:1}). Therefore this
degeneracy issue is serious and needs urgent attention.
We lay down the methodology to obtain a unique two temperature solution using the principles of second law of thermodynamics.
We stated that the solution with the highest entropy near the horizon is the correct solution. 
{Since the proposed correct transonic solution is the one with the highest entropy, therefore it is warranted that these solutions
should be stable for the relevant boundary conditions. Infact, the collective wisdom of the community on accretion solutions
expect, that close to the
horizon accretion should be transonic. 
However, for a given set of accretion disc parameters like ${\dot M}$, $E$ and $\lambda$, we do have a large number of transonic
two temperature solutions, and the question of stability of the solutions arise. In Sect. \ref{sec:stable}, we
showed that the gradient of entropy of the flow with the proton temperature i. e., $d\mdotin/d\tpi$, is such that, it tends to push the
solution towards the temperature corresponding to the highest entropy solution. In other words, if our proposed solution is
perturbed, then $d\mdotin/d\tpi$ would automatically try to restore the solution to the one corresponding to the
highest entropy.}

Once we establish the method to obtain the unique two-temperature solution in rotating disc, we investigated
the effect of various disc parameters on two temperature accretion solutions. We obtained all possible solutions depending on
$E$ and $\lambda$ for a given ${\dot M}$ and $\mbh$ (Fig. \ref{fig:7}) and in addition we also plot the emitted spectrum
(Fig. \ref{fig:8}).
This also shows ${\dot M}$ or $\mbh$ do not alone determine the emitted spectrum or even the luminosity. Depending
on $E$ and $\lambda$, the solution changes and so does the spectrum and luminosity. 
The constants of motion are uniquely linked to the obtained spectrum.

There are indeed shocked accretion solutions even in the two-temperature regime. The shocked solutions
are more luminous, because in the post-shock region inverse-Comptonization becomes effective, the intensity
of the power-law photons increases, compared to a shock-free solution (Fig. \ref{fig:9}). We also showed that accretion flow can
harbour steady shocks in a small but significant patch of the energy angular momentum
parameter space (see, Fig. \ref{fig:10}). However,
in general, the shock strength in two temperature flow is lesser than that in a one-temperature flow.
Moreover, we did not find any particular spectral signature of the presence of shock, only that the
shocked solution is more luminous than shock-free ones. But, it has been found in cases of low accretion rate flows, where weak
bremsstrahlung feature is visible (in the high frequency end of the spectrum) in a shock-free solution, this feature disappears in
a shocked solution (see Fig.~\ref{fig:8}c1,c2).

Radiative properties of a BH system, depends on $\beta$ and $\betah$ along with $E$, $\lambda$, ${\dot M}$ and $\mbh$. For low values of
$\beta$ and $\betah$, the radiative efficiency is around few percent, but for higher values,
the efficiency can easily cross ten percent, even for the same accretion rate and mass of the BH. We also showed that the spectra becomes
broadband if the mass of the central BH considered is higher, it also becomes more luminous but the spectral index remains roughly the
same.
While with the increase in accretion rate of the BH, the bandwidth of spectra remains the same, while the luminosity and the spectral
index
changes significantly.

{We did not consider pair production from the radiation of the accretion disc. Neither did we consider particle production
due to high energy interaction of the protons. We showed from a posteriori calculations, that pair production is negligible,
and therefore the contribution of the pair annihilation emission in the total spectrum of the disc is also negligible. Similarly,
we showed, with the help of a posteriori estimate that the gamma-ray production from pion decay is also negligible.}

In this work we did not consider viscosity of the flow, but invoking our results from our previous works on
viscous accretion solution in the single temperature regime (and also the Appendix \ref{app:1}), we argued
that since the angular momentum is almost constant in the inner part of the disc, and that the viscous heating is much weaker than
the magnetic dissipation, we neglected viscosity to ease our computation. Considering the single
temperature viscous flow as the representative case, we may conclude that, there will not be any qualitatively change by considering
viscosity, although quantitative effect cannot be ruled out. Infact since we showed magnetic dissipation to be a very efficient
heating process, we could incorporate most heating processes by tuning the parameter $\betah$.

To conclude, {it is absolutely necessary to obtain} unique transonic two-temperature solution.
To interpret observations, it is compulsory that
the hydrodynamics of the system is properly handled. Selecting any arbitrary solution would mislead us. We allowed the second law
of thermodynamics to dictate and select the solution, without taking recourse to any assumptions, such that consistency is
maintained.
In addition to spherical flow \citep{sc19}, accretion disc solution also correspond to the highest entropy
solution.

\section*{Acknowledgments} 
The authors acknowledge the anonymous referee for helpful suggestions which improved the quality of the paper. 
SS acknowledges Mr. Kuldeep Singh, for the help in python plotting.

\begin{appendix}
\section{Effect of viscosity in the system :}
\label{app:1}
In this appendix we discuss the effect of viscosity as (1) a mechanism to remove angular momentum outwards and (2) a source of
heating in the system. Handling of viscosity is not trivial and applying it to two-temperature flows might further complicate the
scenario and divert us from the question at hand, which is to find the unique transonic two-temperature accretion solutions for
rotating flows.
Presently, we recall the physics of viscosity in a relativistic but a single-temperature disc like \citet{ck16}
and show that near the horizon viscosity would have marginal effect. And in the outer region it would have a more significant
effect, but since that region contributes less in the spectrum, so for our purpose, we can neglect it without compromising on the
qualitative aspect of the present work.
In viscous one-temperature flows we have azimuthal component of radial-momentum
equation, the integrated form of which is:
\begin{equation}
-\rho u^r(L-L_0)=2\eta \sigma^r_\phi
\end{equation}
where, $L_0$ is the bulk angular momentum at the horizon. $L$ is the local bulk angular momentum expressed as $L=h u_\phi=hl$.
The specific angular momentum is defined as $\lambda=-u_\phi/u_t=-l/u_t$. The dynamical viscosity coefficient is
$\eta_{\rm vis}=\rho h \nu_{\rm vis}$.
Here, $\nu_{\rm vis}=\alpha_{\rm vis} a r (1-v^2)^2$, is the kinematic viscosity, $\alpha_{\rm vis}$ being the \citeauthor{ss73} viscosity
parameter.  $\sigma^r_\phi$ is the $r-\phi$ component of the shear tensor which has been evaluated using the expression given in
\citet{pa97,ck16}. We abbreviate this form of viscosity as PA. Using the same procedure followed in \citet{ck16}, we find solutions for
$E=1.0005, \alpha_{\rm vis}=0.01, \lambda_0=2.60, \dot{M}=0.01\medd$ and $\mbh=10\msol$. Also, we investigate another case of
one-temperature flows using the same set of parameters but assuming a form of viscosity which is generally followed in
non-relativistic accretion disc equations and is given by
$t_{r\phi}=-2\eta_{\rm vis} \sigma_{r\phi}=-\alpha P$. We abbreviate it as SS form of viscosity.
We plot in Fig. \ref{fig:a1}a, the angular momentum distribution of the system as a
function of radius where each curve represent PA form of viscosity (green, solid), and SS viscosity (magenta, dotted).
It is evident from the figure that angular momentum has been transported outwards (for both the cases) due to the
presence of  viscosity. But within $\sim 1000\rg$, angular momentum is almost constant, similar to the case of inviscid flows. The SS
form of viscosity is weaker so it is less efficient in removing angular momentum and hence angular momentum remains almost constant
$\lesssim 3 \times 10^4 \rg$. So, neglecting viscosity within these regions does not affect the system qualitatively. Viscosity in
addition to removing angular momentum, also heats up the system. We plot in Fig.~\ref{fig:a1}b the heat dissipated due to presence of
PA form of viscosity (green, solid) and SS form of viscosity (magenta, dotted). We see that using SS
viscosity is inefficient in heating up the system and is always $3$ orders of magnitude less than PA viscosity.
Very far away from the BH, it is $9$ orders of magnitude less than the latter. We also compare the heating due to magnetic dissipation
($Q_{\rm B}$, blue, dashed), which is the source of heating in this work of two-temperature inviscid flows. In Fig. \ref{fig:a1}b,
$Q_{\rm B}$ has
been calculated using the velocity, temperature and pressure of the corresponding one-temperature flow and using $\betad=0.02$. We see
that $Q_{\rm B}$ is an efficient source of heating in the system. It always supersedes the heating due to presence of GR viscosity except in a
very small region near the BH, where it becomes comparable. Thus, our assumption of taking $Q_{\rm B}$ as a source of heating in the absence of
viscosity, suffice our problem. If viscosity would have been present then the total heating would not have changed much except in
a very narrow region.

\begin {figure}[h]
\begin{center}
 \includegraphics[height=5cm,width=11.cm,trim={0.cm 10.5cm 0.cm 1.5cm},clip]{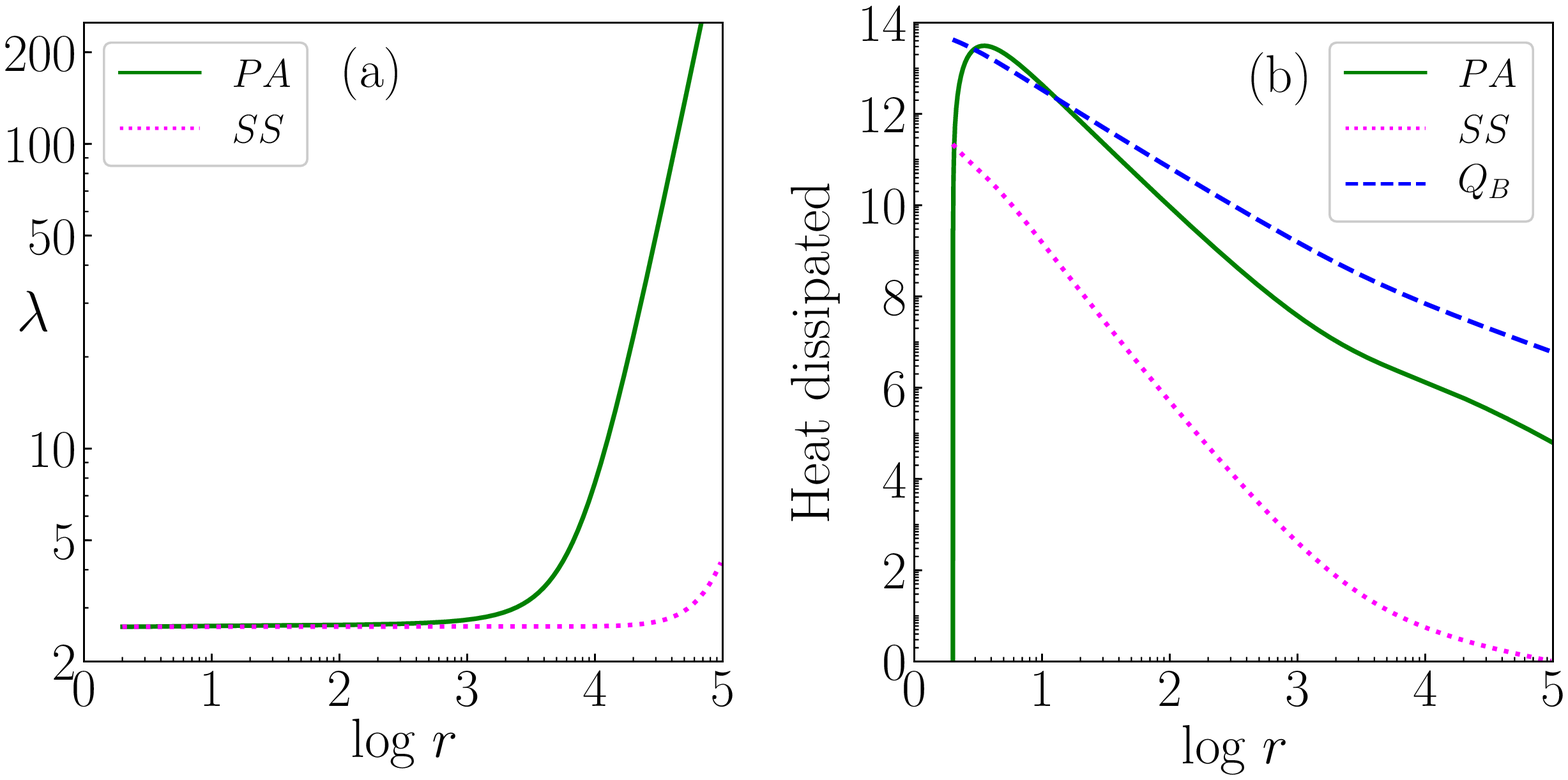}
 \caption{We present (a) distribution of specific angular momentum ($\lambda$) and (b) heating in the system as a function of
 radius (log $r$), when viscosity assumed is relativistic (green, solid) and Newtonian (magenta, dotted). In panel (b) we also
 compare the heating due to magnetic dissipation, $Q_B$ (blue, dashed), assuming $\betad=0.02$. The flow parameters are
 $E=1.0005, \alpha_{\rm vis}=0.01, \lambda_0=2.60, \dot{M}=0.01\medd$ and $\mbh=10\msol$.}
\label{fig:a1}
 \end{center}
\end{figure}

\section{{Estimation of electron-positron pair production in advective two-temperature accretion disc:}}
\label{app:pair}
{In this section we would estimate how much pair can be produced using the two-tmperature accretion solution as
the back ground solution. There are three processes which could lead to pair (electron and positron) production in accretion
discs namely: photon-particle (electron, positron or proton) interaction, particle-particle interaction and photon-photon
interaction.
\cite{s82a,s82b,s84} gave a detailed analysis of the effect of electron-positron pairs present in relativistic and mildly
relativistic plasmas. From these papers it was apparent that photon-photon interaction is the dominant process responsible for
the generation of pairs in accretion discs around BH. The reason behind particle-particle interactions and photon particle
interactions to be of less importance is their excessively small reaction cross-sections, which are of the order of $1/137$
(value of fine structure constant) and $(1/137)^2$ respectively.} 
{The photon-photon pair production process has a threshold condition, which is $E_1 E_2 (1-cos\theta_{pp} ) \geq 2(m_e c^2)^2 $, 
where $E_1$ and $E_2$ are the energies of the photons and $\theta_{pp}$ is the angle between these two photons.
The advective two-temperature accretion disc solution
take into account synchrotron emission which can produce ample amount of soft photons. These photons are too soft to satisfy the 
criterion for pair production \citep{esin99}. But these photons after interacting with high energy electrons can get upscattered 
to high energies, contributing to pair production. 
Also, the bremsstrahlung emission process produce ample amount of hard photons.  Thus at any particular radius we can 
assume the radiation field to be made of a flat bremsstrahlung spectrum which is flat with a high energy cut-off and a 
Comptonization spectrum which is the sum of cut-off power law and the Wien tail \citep{gs67,z85,esin99}. We used the formula given 
by \citet[][see Eq.~B1 of the paper]{s84} to compute the rate of photon photon pair production from two power law (PL) photon
distribution with an 
exponentially cut-off, Wien photon interaction (W-W) and power law photons with Wien photons (PL-W). The pair density
is estimated
a posteriori, by using the temperature profile and velocity profile of the two-temperature solution of this paper. The positron
number density is computed from
\begin{equation}
n_{e^+}=\frac{1}{rH}\int(S^+-S^-)rHdr.
\label{eq:pair}
\end{equation}
Here, $n_{e^+}$ is the positron number density, $S^{\pm}$ are the source and the sink terms or pair production and annihilation rate,
respectively.
$S^\pm$ rates are adopted from \citet[][respectively]{s84,s82a}. Since the accretion disc studied in this paper is
composed of $\ep$ fluid,
so we first integrate Eq. \ref{eq:pair} with only $S^+$ term to compute the maximum possible $n_{e^+}$ produced. With this
distribution of $n_{e^+}$, we compute annihilation rate. We iterate few times till the solution converges.
To estimate the production of electron-positron pairs (\ie $~\el$), we chose two sets of accretion disc parameters presented in the manuscript,
since the spectrum for this disc parameters is hard and is possible to obtain significant hard photons which satisfy
pair-production threshold conditon mentioned above.
We compute pair production for the case of $\betad=0.013$ of Fig.~\ref{fig:12}, and present in 
Figs.~\ref{fig:b1}a1, a2 the same two cases $\dot{M}=1.0, ~\beta=0.2$ and (b1, b2) $\dot{M}=1.5, ~\beta=0.15$. 
In the upper panels (a1, b1), we compare the proton number density $\np$ (black, solid), estimated positron number density
$n_{e^+}$ when annihilation rate is ignored (blue, dotted) and the ones where both production and annihilation rates
are considered (green, dashed). The blue curve is the maximum possible positrons that can be produced in the disc
and $n_{e^+} \ll \np$. In the bottom panels (a2, b2), we plot the corresponding emissivity $Q_{\rm ann}$
obtained due to the annihilation of pairs (red, dotted) and compared that with the total emissivity $Q_{\rm tot}$ (brown, solid).
The estimated number density of positrons is negligible and the contribution to the total emissivity is negligibly small
compared to the total emissivity obtained from radiative processes like, synchrotron, bremsstrahlung and Comptonization.
This a posteriori estimation justifies our assumption of not considering pair production in the present work.}

\begin{figure}[h!!!!]
\begin{center}
\includegraphics[width=12cm,height=8cm,trim={0.2cm 1.0cm 0.2cm 7.0cm},clip] {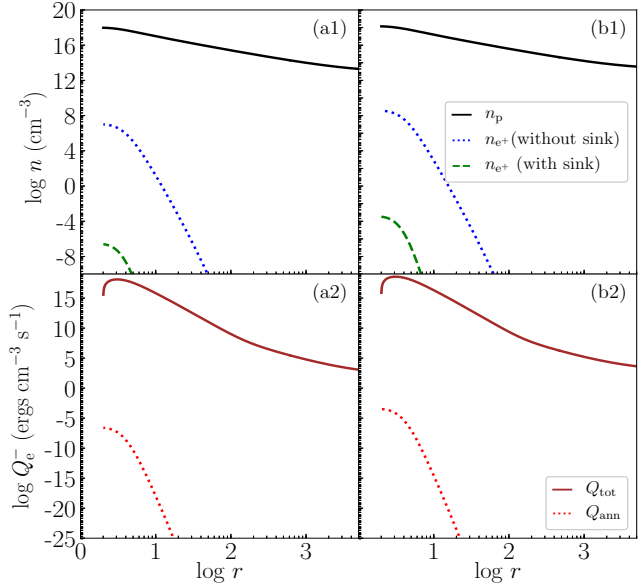}
\caption{\label{fig:b1} {(a1, b1)
Comparison of number density of protons $\np$ (black, solid), positron number densities
$n_{e^+}$ (without annihilation, blue, dotted) and with both production and annihilation rates (green, dashed). (a2, b2)
Comparison of emissivities of the total radiative cooling $Q_{\rm tot}$ and emissivities due to annihilation of pairs
$Q_{\rm ann}$. For two sets of accretion disc parameters,
(a1, a2) $\dot{M}=1.0, ~\beta=0.2$ 
and (b) $\dot{M}=1.5, ~\beta=0.15$. The other parameters are $\betad=0.013$, $E=1.001$, $\lambda=2.61$ and $\mbh=10\msol$.}}
\vskip -0.95cm
\end{center}
\end{figure}

\section{{Estimation of the gamma-ray emission by pion interaction:}}
\label{app:pion}
{
In this section we discuss whether pion production lead to viable amount of cooling in the disc and have any observational
signature.
The reactions leading to pion ($\pi^\pm,~\pi^0$) production by proton-proton interactions are as given below \citep{el80}}:
\begin{align}
&p+p \rightarrow p+n+\pi^+ \nonumber \\
&p+p \rightarrow p+p+\pi^0 \nonumber \\
&p+p \rightarrow d+\pi^+ \nonumber \\
\end{align}
{The threshold temperature for these reactions is $290$ MeV. For negative pions temperatures of $>2$ GeV are required.
Thus, it is can be assumed that negligible $\pi^-$ will be present in the disc. The $\pi^0$ further decay
into gamma-ray photons \citep{ks79}:}
\begin{equation}
 \pi^0 \rightarrow 2\gamma_{\rm ph}
 \label{eq:piprod1}
\end{equation}
{and, $\pi^+$ decays into muon and muon neutrinos, which further decays into electron neutrinos, muon anti-neutrinos and
positrons, respectively.}
\begin{align}
& \pi^+ \rightarrow \mu^+ +\nu_\mu \nonumber \\
& \mu^+ \rightarrow \nu_e + e^+ + \bar{\nu}_\mu
\end{align}
{We restrict our study to neutral pions $\pi^0$ since we are interested to study the gamma ray emissivities obtained from an
accreting
Schwarzschild BH. The rate of $\pi^0$ production is given by (in units of cm$^{-3}$ s$^{-1}$):}
\begin{equation}
{\cal{R}}_{\pi^0}=\frac{n^2}{2} <\bar {\sigma} \bar{ v}>_{\pi^0}.
\end{equation}
{Here,  $<\bar {\sigma} \bar{ v}>_{\pi^0}$ (units of cm$^{3}$ s$^{-1}$) is the velocity weighted cross-section, which was
evaluated for $\pi^0$ by \citeauthor{dcw74} in 1974, assuming experimental cross sections for pion production and a relativistic
Maxwell-Boltzmann distribution for protons. This was further investigated by \citet{w76} and \citet{ks79} who computed
$<\bar {\sigma} \bar{ v}>$ for $\pi^0$ as well as for $\pi^+$. This function is strongly dependent on the proton temperature.
In 1986 \citeauthor{cmt86} obtained a best fit to these curves, expression of which is given in Eqs.10 of their paper,
and the same form is used to compute the emissivity of $\gamma$-rays. It is clear from Eq.~\ref{eq:piprod1} that each
$\pi^0$ decays into two photons, therefore the number of photons produced per unit time per unit volume is
${n^2} <\bar {\sigma} \bar{ v}>_{\pi^0}$.
We analyzed a posteriori the total gamma-ray luminosity as measured by an observer at infinity using the methodology adopted in
\citet{cmt84}.}

{We checked the production of $\gamma$ rays by varying the accretion rate of the system from $\dot{M}=0.01$ (red, dotted) to
$0.10$ (green, dashed) and $1.0$ (blue, solid) for a $10\msol$ BH (see Figs.~\ref{fig:c1}a1, a2). Since this emission is crucially
dependent on the proton temperature we plot log$\tp$  as a function of log $r$ in Fig. \ref{fig:c1}a1 for the different
accretion
rates.
The set of disc parameters used are $\lambda=2.61$ and $E=1.0007$.   For ${\dot M}=0.01,~\&~0.1$, the accretion solution
do not undergo shock transition (see, Fig.~\ref{fig:9}) and the $\tp$ distribution of the global accretion solution are similar.
But for the same set of $E$ and $\lambda$ and ${\dot M}=1.0$, there is a stable accretion shock and the $\tp$ jumps at the shock
location.
The corresponding spectra (Fig.~\ref{fig:c1}a2) for the three values of ${\dot M}$ show marked difference, 
where the shocked accretion solution is more luminous and the spectrum is harder (solid, blue), and becomes less luminous and
softer for lower ${\dot M}$. The gamma-ray emission (calculated a posteriori) is represented in grey colour. 
As a result of $\pi^0$ decay, the contribution in the high energy regime increases.
Quantitatively, the gamma-ray luminosity for different accretion rates
are related by the following relation $L_{\gamma_{\rm ph}} (\dot{M}=1.0) \simeq 200 L_{\gamma_{\rm ph}} (\dot{M}=0.1)$
and $L_{\gamma_{\rm ph}} (\dot{M}=0.1)\simeq 100 L_{\gamma_{\rm ph}} (\dot{M}=0.01)$.
Although $L_{\gamma_{\rm ph}} (\dot{M}=1.0)$ is much higher than that compared to lower ${\dot M}$s but compared to the
overall luminosity for each ${\dot M}$, $L_{\gamma_{\rm ph}}$ is pitiably low.
We chose ${\dot M}=0.1$ and the same values of $E$ and $\lambda$ and then studied the gamma-ray production for accretion discs
onto different masses of central BH. We plot the proton temperature distribution
$\tp$ vs $r$ in log-log scale and the corresponding spectra in Figs.~\ref{fig:c1}b1, b2, for central BH masses $\mbh=10^2$ (blue, solid),
$10^4$ (green, dashed) and $10^6$ (red, dotted). $\mbh$ affects the system quantitatively but not qualitatively. While
the luminosity increases and spectra becomes more broad band with the increase in $\mbh$, but the efficiency of
 gamma-ray emission ($L_{\gamma_{\rm ph}}/(\dot{M} c^2)$) remains almost same $\sim 10^{-8}$.
 In both the cases we examined here (change in accretion rate
and mass of BH), the fractional change in luminosity is always $< 10^{-5}$. This analysis justifies our assumption of neglecting
pion production leading to emission of $\gamma$-rays.}

\begin {figure}[h]
\begin{center}
 \includegraphics[height=10cm,width=12cm,trim={1.cm 3.cm 1.3cm 0.8cm},clip]{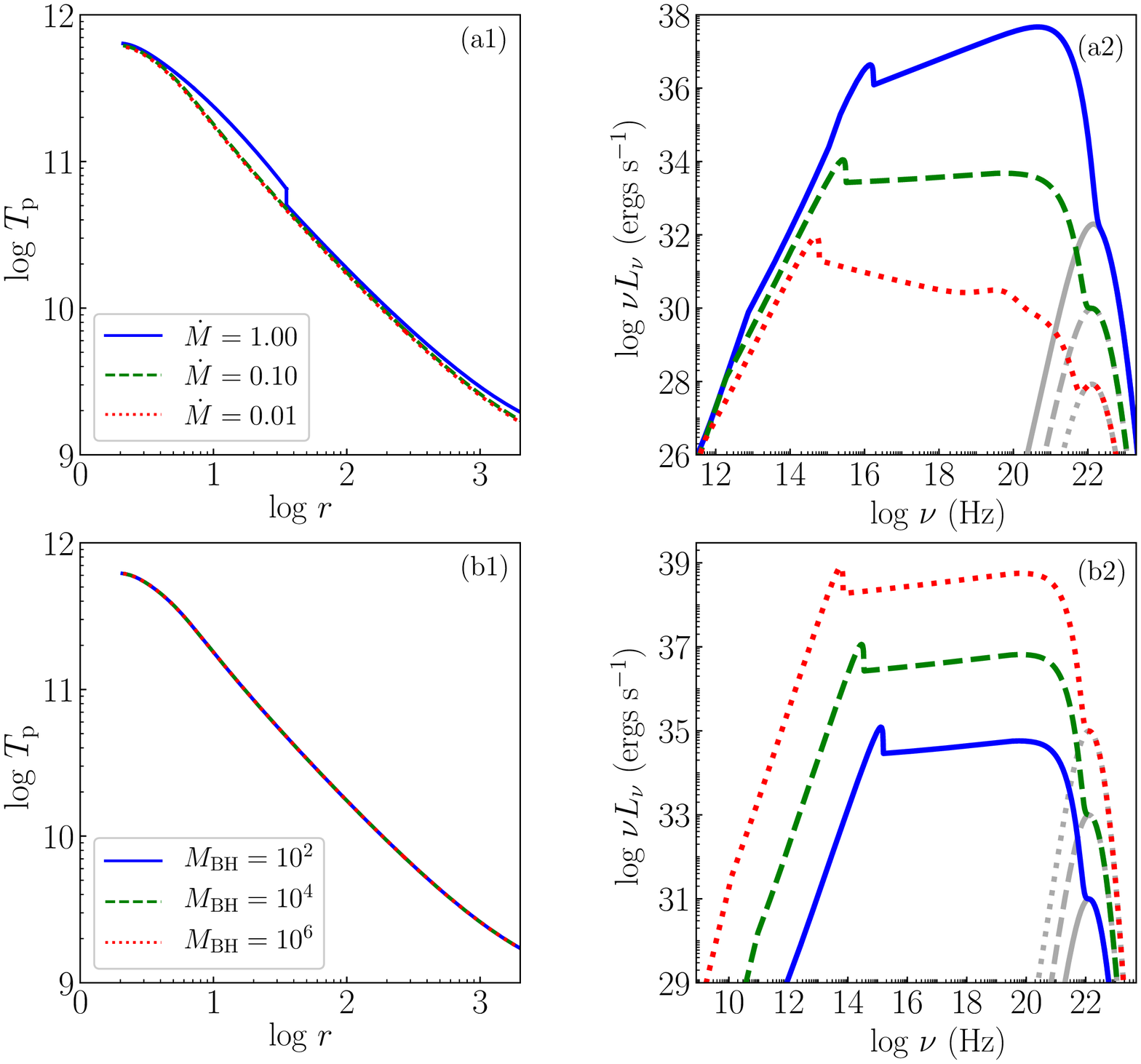}
 \caption{{(a1, a2) Dependence on accretion rate $\dot{M}=0.01$ (red, dotted), $0.1$ (green, dashed)
 and $1.0$ (blue, solid). (a1) log~$\tp$ as a function of log~$r$ and (a2) log~$\nu L_\nu$ with log~$\nu$.
 (b1, b2) Dependence on mass of BH, $\mbh=10^2$ (blue, solid), $10^4$ (green, dashed), $10^6$ (red, dotted) in
 production of $\gamma$ rays. (b1) log~$\tp$ as a function of log~$r$ and in (b2) the corresponding spectra are plotted.
 The
 gamma ray emission is presented in grey.}
}
\vskip -0.75cm
\label{fig:c1}
 \end{center}
\end{figure}

\end{appendix}

\end{document}